\documentclass[12pt,preprint]{aastex}
\def\h1{$^1$H}
\def\o16{$^{16}$O}
\def\c12{$^{12}$C}
\def\n14{$^{14}$N}
\def\he4{$^4$He}
\everymath{\rm}
\everydisplay{\sf}
\received{August 1, 2004}
\revised{August 31, 2004}
\accepted{September 1, 2004}
\slugcomment{To appear in {\it The Astrophysical Journal}}
\shortauthors{Nava \& Henry}
\shorttitle{N and O At Low Metallicity}
\begin{document}
\title{The N/O Plateau of Blue Compact Galaxies: Monte Carlo Simulations of the Observed Scatter}

\author {R.B.C. Henry, A. Nava}

\affil{Homer L. Dodge Department of Physics \& Astronomy, University of Oklahoma,
Norman, OK 73019; nava,henry@nhn.ou.edu}

\and

\author {Jason X. Prochaska}

\affil{Lick Observatory, University of California, 1156 High Street, Santa Cruz, 
CA 95064; xavier@ucolick.org}

\begin{abstract}

Chemical evolution models and Monte Carlo simulation techniques have been combined for the first time to study the distribution of blue compact galaxies on the N/O plateau. Each simulation comprises 70 individual chemical evolution models. For each model, input parameters relating to a galaxy's star formation history (bursting or continuous star formation, star formation efficiency), galaxy age, and outflow rate are chosen randomly from ranges predetermined to be relevant. Predicted abundance ratios from each simulation are collectively overplotted onto the data to test its viability. We present our results both with and without observational scatter applied to the model points. Our study shows that most trial combinations of input parameters, including a simulation comprising only simple models with instantaneous recycling, are successful in reproducing the observed morphology of the N/O plateau once observational scatter is added. Therefore simulations which include delay of nitrogen injection are no longer favored over those which propose that most nitrogen is produced by massive stars, if only the plateau morphology is used as the principal constraint. The one scenario which clearly cannot explain plateau morphology is one in which galaxy ages are allowed to range below 250~Myr. We conclude that the present data for the N/O plateau are insufficient by themselves for identifying the portion of the stellar mass spectrum most responsible for cosmic nitrogen production.

\end{abstract}



\section{INTRODUCTION}

Dwarf irregular galaxies (dIs) and their bursting derivatives, blue compact 
galaxies (BCGs), are small, low mass systems which are relatively metal-poor 
(see Kunth \& {\"O}stlin 2000 for an excellent review of metal-poor galaxies). From the standpoint of galactic chemical evolution, one of the valuable 
features of dIs and BCGs is that they serve as excellent probes of elemental buildup at low 
metallicities, since the emission spectra from the photoionized gas associated with them enables the abundances of oxygen and numerous other elements to be readily determined. Using oxygen as a metallicity tracer, 12+log(O/H) is generally observed to range between 7.1 and 8.1 for these objects, where solar oxygen is 8.66 (Asplund, Grevesse, \& Sauval 2005), and oxygen in the LMC and SMC are 8.4 (Garnett 1999) and 8.15 (Peimbert et al. 2000), respectively. The lowest metallicity dIs include I~Zw~18, with a value of 7.17 (Garnett et al. 1995; see also the studies by Izotov \& Thuan 1999), UGCA~292, with a value of 7.32 (van~Zee 2000; van~Zee \& Haynes 2006), and SBS~0335-052W, with a value of 7.13 (Papaderos et al. 2006). All three are BCGs. 

When one measures N and O abundances in BCGs and then plots log(N/O) versus 12+log(O/H) 
[henceforth referred to as N/O and O/H, respectively], it is found that these 
objects form a plateau, consistent with the notion that stellar N production over the BCG metallicity range is primary (see the Appendix for a review of nitrogen synthesis basics). In the past, much of the observed spread in N/O values along the plateau has implicitly been assumed to be real and not simply due to measurement uncertainties (see Henry, Edmunds, \& K{\"o}ppen 1999, for example). 

\begin{figure}
\centering
\figurenum{1}
\includegraphics[width=12cm,angle=270]{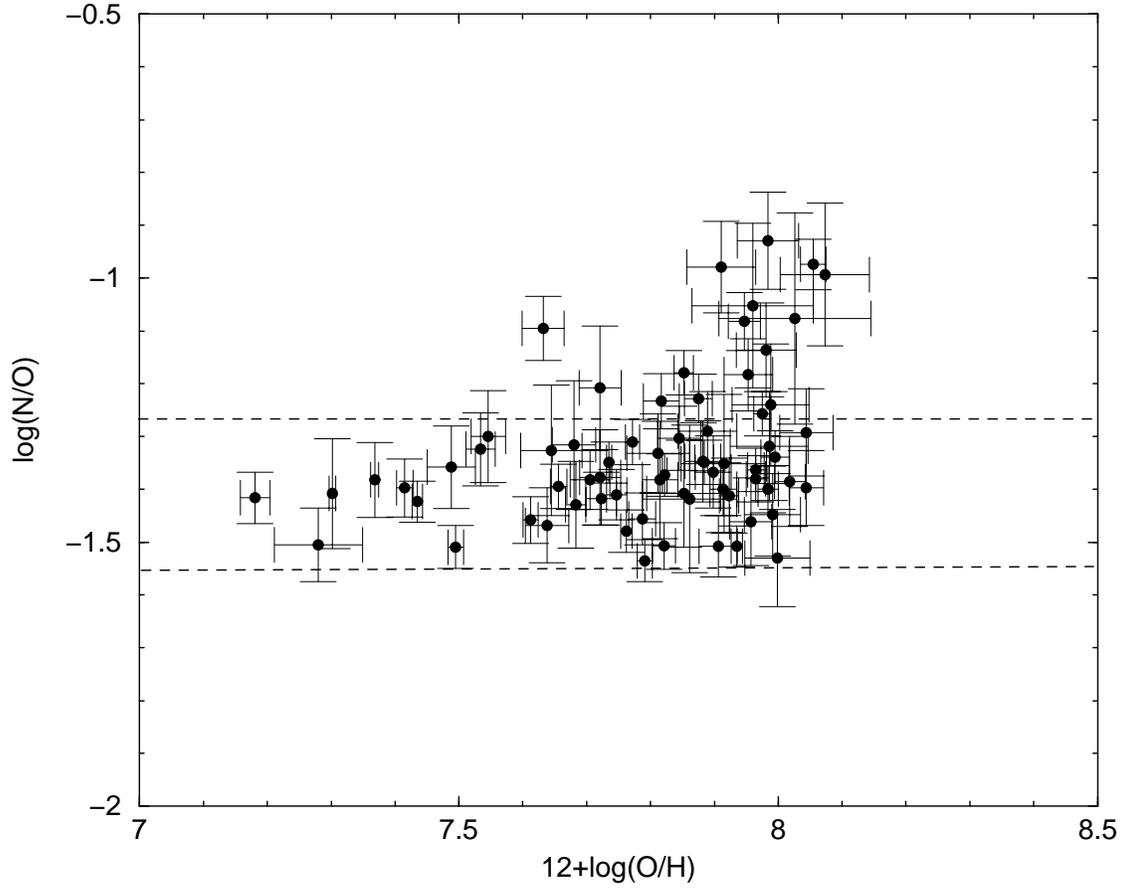}
\caption{Plot of log(N/O) vs. 12+log(O/H) for the sample of blue compact galaxies 
studied by Nava et al. (2006). The horizontal dashed lines show the vertical boundaries for
the N/O plateau as defined by Nava et al.}
\label{aida_final}
\end{figure}

However, Nava et al. (2006) have recently rederived N and O abundances from published line strengths in order to pin down uncertainties as carefully as possible and then to quantify any natural component of the scatter observed for N/O. They employed direct abundance determining techniques but supplemented them with photoionization models using CLOUDY (Ferland 1998) and input stellar spectra tailored appropriately to match the metallicity range and N/O values found along the plateau. These input spectra were calculated using the stellar atmosphere code PHOENIX (Hauschildt \& Baron 1999, 2004). Their abundance results are plotted in Fig.~\ref{aida_final}, which shows N/O vs. O/H for each of their points, along with uncertainties. 

Based upon their abundance determinations, Nava et al. found that BCGs fall into two groups according to their N/O values, where the larger group exhibits a near-Gaussian distribution in this quantity, while objects in a smaller group possess significantly higher N/O levels. They strictly defined the N/O plateau as the region in Fig.~\ref{aida_final} stretching horizontally between 7.1$\le$O/H$\le$8.1 and vertically between -1.54$\le$N/O$\le$-1.27, excluding objects associated with the high N/O tail. Plateau BCGs were found to have a mean value and error of -1.43 $\pm$.0084/.0085. Nava et al. then carried out a $\chi$-square analysis of the Gaussian portion of the N/O distribution to demonstrate quantitatively for the first time that most, if not all, of the observed spread in N/O is not intrinsic or real but is primarily the result of uncertainties in line measurments.

The nature of the scatter on the N/O plateau bears heavily on the problem of pinning down the precise stellar mass ranges which are responsible for the production of nitrogen in the universe. We assume that future observations with large telescopes and reduced uncertainties will help confirm or deny the results of Nava et al. However, in order to move forward on the theoretical side of the issue, we first assume here that the observed scatter is intrinsic and then proceed by combining numerical chemical evolution models with Monte Carlo techniques in order to see if any of the popular scenarios for producing variations in N/O can be ruled out by the data. We then adjust the original simulation predictions by smearing their output points to imitate observational uncertainty and further compare with the observations.

In order to appreciate the problem which intrinsic scatter in N/O presents, let us first apply a simple model analysis to the situation. The simple model of chemical evolution assumes a closed box, instantaneous recycling, and relates the mass fraction of an element Z to the product of the stellar yield y and a logarithmic term involving the inverse of the gas fraction $\mu$, i.e. $Z=y\ln(1/\mu)$. Thus, assuming that the O yield is metallicity independent, the horizontal spread in O/H along the plateau arises because sample objects differ in the extent to which they have formed stars, i.e. those more evolved have smaller gas fractions. On the other hand, for an element-to-element ratio such as N/O, the value is 
set by the ratio of the respective integrated stellar yields of N and O, since the 
gas fraction terms cancel. So, the simple model predicts that when yields are independent of metallicity, objects of different metallicity should nevertheless have the same N/O ratio and thus form a very narrow track roughly parallel to the horizontal axis with essentially no intrinsic scatter in N/O.

Given the simple model prediction, then, there are basically two options for explaining the natural spread in N/O along the plateau. The first one is the delayed release of N relative to that of O following an episode of star formation when N is assumed to be principally produced by intermediate-mass stars (IMS), objects having masses below 8~M$_{\odot}$
which evolve slower and expel their products later than the massive stars responsible for the O production. If such delay is relevant, then a sample of galaxies comprising objects at various stages following a star burst should exhibit a range in N/O values\footnote{Note that if both N and O are forged by massive stars, then the delay question becomes irrelevant.}. Chemical evolution models which use variations in star formation history, including continuous and bursting star formation processes, fall within this class. 

The second option is the alteration of the effective yield, i.e. the yield value inferred from the simple model when
the observed elemental mass fraction and gas fraction are considered. If the effective yield of either N or O varies from galaxy to galaxy, then scatter in N/O will result. Chemical evolution models which employ differing infall and/or outflow rates among galaxies follow this option.

Identifying the cause of the N/O spread in
low metallicity galaxies is a pressing issue intimately related to the question of the chemical evolution of N and O.  Various aspects
of this problem, along with the
closely-related one concerning the identification of the stellar origin of nitrogen,
have been addressed empirically by Edmunds \& Pagel (1978), Lequeux et al. 
(1979; 1981), Serrano \& Peimbert (1983), Garnett (1990), Alloin et al. (1979), 
Vila-Costas \& Edmunds (1993), Thuan, Izotov, \& Lipovetsky (1995), Izotov \& 
Thuan (1999), Pilyugin, Thuan, \& V{\'i}lchez (2003), Pilyugin, Contini, \& 
V{\'i}lchez (2004), Thurston, Edmunds, \& Henry (1996), Kobulnicky \& Skillman 
(1996, 1997, 1998), Kobulnicky et al. (1997) van~Zee, Salzer, \& Haynes (1998), van Zee et al. (1998), van Zee \& Haynes (2006), 
and Izotov et al. (2004). At the 
same time, complementary theoretical work has been carried out by Matteucci \& 
Chiosi (1983), Matteucci \& Tosi (1985), Matteucci (1986a,b), Pilyugin (1992; 
1993), Marconi, Matteucci, \& Tosi (1994), Bradamante, Matteucci, \& D'Ercole (1998),
 Henry, Edmunds, \& K{\"o}ppen (2000; hereafter HEK), Chiappini, 
Romano, \& Matteucci (2003), Chiappini, Matteucci, \& Meynet (2003),
Lanfranchi \& Matteucci (2003), Chiappini, Matteucci, \& Ballero (2005),
and K{\"o}ppen \& Hensler (2005). 

Modern chemical evolution models employing current yields have been used to interpret the most up-to-date abundance data on BCGs. Models purporting to explain N/O scatter 
based on star formation history  
range from those preferring continuous star formation (HEK) to those supporting bursting star formation (Bradamante et al. 1998; Chiappini, Romano, \& Matteucci (2003); and Lanfranchi \& Matteucci 2003). Bradamante et al. also stress the importance of a variable star formation efficiency (SFE) in producing the scatter. All of these models primarily explore mechanisms related to the delay in N release. In contrast, models claiming to solve the problem through infall have recently been advanced by K{\"o}ppen \& Hensler (2005), where their approach centers mainly on effective yield changes to explain the observations. Currently, there is no solid consensus on the cause of the scatter. 

{\it The goal of this paper is to examine the potential causes of the vertical scatter along the N/O plateau.} In this study for the first time we combine detailed chemical evolution 
models and Monte Carlo simulation techniques in an attempt to reproduce the morphology of the empirical N/O plateau as defined for purposes of study by Nava et al. (2006). Each Monte Carlo 
simulation comprises numerous chemical evolution models calculated using parallel 
processing, where the values of the input parameters for each model are chosen at 
random from within a realistic range for each. While adopting one set of stellar yields throughout, our study carefully looks at the two principal options for explaining N/O morphology, namely,
time delay and effective yield variations. 

The paper is organized as follows. Section~2 describes the methods of our study, 
including a description of our chemical evolution code as well as the details of 
our Monte Carlo simulations, {\S}3 provides the results and discussion of our 
study, and {\S}4 gives a summary and our conclusions.

\section{METHODS OF ANALYSIS}

\subsection{The General Approach}

Our approach to the analysis of the BCG abundances breaks with the tradition of 
overplotting tracks associated with individual models onto the observational 
plane and then using these tracks to interpret the data, where each track shows the 
evolution of N/O and O/H with time under a single set of initial conditions. 
Such a practice can be misleading, as it often implies an evolutionary 
connection between individual data points while deemphasizing the varied origins 
and initial conditions of the systems represented in the plot. Chiappini, Matteucci, and co-workers in particular have always 
emphasized this last point. Using the usual approach of plotting models directly in the N/O-O/H plane also obscures the time dimension, and so the effects of delay are difficult to study.

Here, we employ Monte Carlo simulations in an attempt to reproduce the point 
distribution of the N/O plateau as defined by Nava et al, making the assumption for now that {\it all} of the scatter described by these authors is intrinsic. In so 
doing, we test the importance of the two options described above, i.e. delay in the release of N into the ISM, as well as effective yield variations in explaining the 
vertical scatter in N/O.

In the simple model picture presented in the previous section, it is assumed 
that the current stellar population synthesizes and expels both N and O 
simultaneously (the instantaneous recycling approximation). However, there is 
both good evidence as well as strong theoretical implications (Chiappini, Romano, \& Matteucci 2003; Henry 2004)
that N is produced mainly by intermediate mass stars (IMS), while 
faster-evolving massive stars are responsible for the production of O. Thus, a 
variation of N/O may result because observations of BCGs sample a population of 
objects in different stages of N release. Within this scenario, those objects exhibiting
low N/O values have experienced O and N release by their massive star component, while N has
yet to be expelled by IMS, because not enough time has elapsed since the last episode of star formation.
On the other hand, in objects with N/O values roughly equal to the ratio of integrated N and O yields, N ejection by IMS has occurred; galaxies with intermediate values are currently experiencing N release and so are evolving toward higher
levels of N/O. This picture has been the 
most popular one lately for explaining the vertical scatter in N/O. In 
particular, recent studies by HEK, Chiappini, Romano, \& Matteucci (2003), and Pilyugin et al. (2003) 
have proposed that effects of delay coupled with those of system age primarily 
explain the scatter.

In studying the relevance of delay we treat both the cases of continuous and 
bursting star formation. In the former, star formation continues without stopping after 
it begins, although its rate may vary as the gas fraction declines. We contrast 
this situation with that of bursting in which star formation is turned on and 
off over prescribed intervals and for specified durations.

The other major process which we explore is outflow as a mechanism for varying the effective yield.
Here we suppose that 
elements are selectively removed from the galactic system as the result of being 
propelled outward in the presence of a relatively weak surface potential. The 
proposed driving mechanism is the energy supplied by Type~II supernovae. Since 
BCGs are small in mass, their surface potentials are likely to be weak, and thus 
ejecta from Type~II events may easily be expelled into the 
intergalactic medium, carrying nuclear products, especially alpha elements such 
as O, with it. At the same time, if N is produced principally in IMS as we assume, where the 
ejection process is much less energetic, N is likely to be retained by the system. As a consequence,
the effective yield of O is reduced, while that of N is 
unchanged, and the N/O ratio rises above the simple model value by an amount 
which can vary from galaxy to galaxy depending on the amount of outflow 
associated with each system.

\subsection{The Chemical Evolution Code}

Our numerical code is an expanded version of the one employed by HEK. It is a one-zone chemical evolution program
which follows
the buildup of the elements H, He, C, N, O, Ne, Si, S, Cl, Ar, and Fe over time. 
Throughout the entire project, we made the assumption that the model galaxy was 
formed by accreting matter before star formation began. This is consistent with the idea advanced
by Skillman et al. (2003) in which star formation in BCGs actually begins several Gyr after
galaxy formation. Thus, each system 
started with an initial mass which changed only if and when material was ejected 
through outflow.

We imagine a generic galactic region with a cross-sectional area of 1~pc$^{-2}$ 
filled with primordial, metal-free gas of mass $M$. As star formation commences, 
the total mass within the region becomes partitioned into interstellar gas of 
mass $g$ and stars of mass $s$ such that
\begin{equation}
M=g+s.\label{mgs}
\end{equation}
If $\psi(t)$ and $e(t)$ are the rates of star formation and stellar ejection, 
respectively, then:
\begin{equation}
\dot{s}=\psi(t)-e(t)
\end{equation}
and
\begin{equation}
\dot{g}=-\psi(t)+e(t).\label{g}
\end{equation}
The interstellar mass of element $x$ in the zone is $gz_x$, whose time
derivative is:
\begin{equation}
\dot{g} z_x + g \dot{z}_x = -z_x(t) \psi(t) + e_x(t),
\end{equation}
where $z_x(t)$ is the mass fraction of $x$ in the
gas, and $e_x(t)$ is
the stellar ejection rate of $x$. Solving for $\dot{z}_x(t)$ yields:
\begin{equation}
\dot{z}_x(t)=\{e_x(t) - e(t)z_x(t)\}g^{-1}.\label{zdot}
\end{equation} 

The rates of mass ejection $e(t)$ and ejection of element $x$, $e_x(t)$, are:
\begin{equation}
e(t)=\int_{m_{\tau_{m}}}^{m_{up}}[m-w(m)]\psi(t-\tau_m)\phi(m)dm\label{e}
\end{equation}
and
\begin{equation}
\begin{array}{lr}
e_x(t)=(1-C)\int_{m_{\tau_{m}}}^{m_{up}}\{[m-w(m)]z_{x}(t-\tau_{m})+mp_x(m,z_{t-
\tau_{m}})\}\psi(t-\tau_m)\phi(m) \ dm\\
\\
+ 
C\int^{M_{BM}}_{M_{Bm}}p_x(m,z_{t-\tau_m})m\phi(m)\int^{.5}_{\mu_{min}}24\mu^2
\psi(t-\tau_{m_2})d\mu\ dm_B. \label{ex}
\end{array}
\end{equation}
In Eqs.~\ref{e} and \ref{ex}, $m$ and $z$ are stellar mass and metallicity, 
respectively, and $m_{\tau_m}$ is the turn-off mass, i.e. the stellar mass whose 
main sequence lifetime corresponds to the current age of the system. This 
quantity was determined using results from Schaller et al. (1992). $m_{up}$ is 
the upper stellar mass limit, taken to be 120~M$_{\sun}$, $w(m)$ is the remnant 
mass corresponding to ZAMS mass $m$ and taken from Yoshii, Tsujimoto, \& Nomoto 
(1996). $p_{x}(z)$ is the stellar yield, i.e. the mass fraction of a star of 
mass $m$ which is converted into element $x$ and ejected, and $\phi(m)$ is the 
initial mass function. Our choice of stellar yields is discussed below.  
In eq.~\ref{ex} the first integral gives the contributions to the ejecta of single stars, 
while ejected masses of C, O, Si, S, Cl, Ar, and Fe by SNIa through binary star 
formation are expressed by the second integral, where our formulation follows 
Matteucci \& Greggio (1986). Assuming that the upper and lower limits for binary 
mass, M$_{BM}$ and M$_{Bm}$, are 16 and 3~M$_{\odot}$, respectively, 
eq.~\ref{ex} splits the contributions from this mass range between single and 
binary stars. The relative contributions are controlled by the parameter $C$ 
which we take to be equal to 0.05 when $3\le M \le 16\ M_{\odot}$ and zero 
otherwise. The variable $\mu$ is the ratio of the secondary star mass $m2$ to 
the total binary mass, $m_B$, where $\mu_{min}=max(1.5,m_{\tau_m},m_B-8)$.

The initial mass function $\phi(m)$ is the normalized Salpeter (1955) relation
\begin{equation}
\phi(m)=\left[\frac{1-\alpha}{m_{up}^{(1-\alpha)}-m_{down}^{(1-\alpha)}}\right] 
m^{-(1+\alpha)},\label{psi}
\end{equation}
where $\alpha$=1.35 and $m_{down}=1$~M$_{\odot}$.
Finally, the star formation rate $\psi(t)$ is given by
\begin{equation}
\psi(t)=\nu M \left(\frac{g}{M}\right)^{1.5}\label{sfr}
\end{equation}
where $\nu$ is the star formation efficiency in Gyr$^{-1}$.

Several of the above equations required adjustments for outflow when the effects of that 
mechanism were being tested. 
Specifically, contributions made by massive stars (M$>$8M$_{\odot}$) 
to the stellar ejection rates, $e(t)$ and $e_x(t)$, were multiplied by a windloss 
parameter, $wl$, i.e. the fraction of stellar ejecta ultimately lost to the intergalactic medium. 
The amount of outflowing matter, including some newly synthesized O and other alpha elements, was also subtracted from the system's total mass $M$. These 
adjustments explicitly affected Eqs.~\ref{mgs}, \ref{g}, \ref{e}, and \ref{ex}.

Our calculations assumed a timestep length of one million years, a value which is
less than the main sequence lifetime of a star with a mass equal to $m_{up}$, or 120~M$_{\odot}$.

 At each time 
point, the increment in $z_x$ was calculated by solving eq.~\ref{zdot} and 
the required subordinate equations \ref{e} and \ref{ex}. This increment was 
then added to the current value and the program advanced to the next time step. 
Finally, the total metallicity at each point was taken as the sum of the mass 
fractions of all elements besides H and He. The unit of time is the Gyr, while 
the mass unit is the solar mass.

Single star yields for all elements but Cl and Ar were taken from Portinari, 
Chiosi, \& Bressan (1998) for massive stars and Marigo (2001) for low and intermediate mass stars. An 
advantage of using these two sets is that they are designed to be complementary. The 
massive star models account for the effects of convective overshooting and 
quiescent mass loss and cover a metallicity range from 0.0004 to 0.05. For 
metallicities below 0.0004, yields were assumed to be equal to their values at 
0.0004, i.e. yields were not extrapolated in metallicity. However, since N 
production is exclusively secondary in massive stars, yields below 0.0004 were 
scaled with metallicity. Since Portinari et al. did not publish Cl and Ar 
yields, we used the results of Woosley \& Weaver (1995). Marigo's computations 
for LIMS account for mass loss, third dredge-up, and hot bottom burning over 
metallicity range of 0.004 to 0.019. Finally, in the case of binary star 
contributions to evolution of C, O, Si, S, Cl, Ar, and Fe through SNIa events, 
we employed the yields of Nomoto et al. (1997), using both the W7 and W70 models 
to account for metallicity effects. We note that the quantities published by Portinari 
et al. represent the sum of the amount present in the star at birth and 
that synthesized by the star.

\subsection{The Monte Carlo Simulations}

In the process of testing the delay and outflow mechanisms, we 
performed roughly 140 separate Monte Carlo simulations, nine of which were ultimately selected for discussion in {\S}~3 below. Each simulation comprised 70 
chemical evolution models calculated simultaneously using parallel processing techniques, where the number 70 was chosen because it roughly matches the number of observed data points. Within a single
simulation certain input parameters remained constant from model to model, while those parameters whose impact on scatter were being tested were chosen randomly using a random number generator which produced numbers of uniform deviatation (all values within a parameter range had an equal probability of being chosen). The range of possible values for a varying parameter was established by the requirement that the predicted O/H range of a simulation had to be roughly consistent with the observed one. The parameters that were chosen at random and allowed to vary under the assumption of continuous star formation were chemical 
age (the elapsed time since the beginning of star formation), star formation efficiency, and the wind loss parameter. 

The aim 
of the Monte Carlo study is to find parameter ranges which produce a set of 
simulated points whose distribution in the plateau region of the N/O-O/H plane best duplicates the 
observations. In the plots presented in {\S}~3 two versions of a simulation are always shown. In one version the points have coordinates exactly as predicted by the models, enabling us to test the impact of varying input parameters while assuming all scatter is intrinsic, i.e. the usual practice. In a second version, the model predicted points have been smeared in order to include the effects of observational uncertainty. The object here is to demonstrate how the original point distribution changes in the presence of observational uncertainty. In this case uncertainties were assumed to have a normal (Gaussian) distribution about the actual point, and so an uncertainty value was determined randomly using a routine which generates random deviates with a Gaussian distribution. In the smearing process, we assumed a standard deviation of 0.073~dex in N/O and 0.031~dex in O/H, where both values are average logarithmic errors taken from Nava et al. The routines employed for smearing were {\it ran3} and {\it gasdev}, both of which can be found in Press et al. (2003).  

\section{RESULTS AND DISCUSSION}

We proceed in this section by discussion all nine of the unsmeared simulations first, as there is academic interest in seeing the effects which the variation of certain parameters have on the outcome in the absence of observational scatter. We then end the section with a discussion of the smeared simulation results. At that point we will see that introducing observational uncertainty into the simulations makes it difficult to discount most of the simulations for explaining the observations.

\subsection{Time Delay}

The effect of time delay in N release can best be demonstrated by calculating a single chemical evolution model for a burst of star formation lasting 100~Myr. The resulting evolution of N/O is shown in Fig.~\ref{singleburst}, where this ratio is plotted as a function of time since the beginning of the burst. The dashed lines indicate the limits of the N/O plateau presented in Fig.~\ref{aida_final} The N/O ratio rises and finally reaches -1.60, the lower limit of the plateau, after about 250~Myr, subsequently leveling off at the value corresponding to the integrated yield ratio, -1.37, at around 400 Myr when N release is complete. Because of the gradual release of N as IMS evolve, a single galaxy moves through the plateau region for about 150~Myr before coming to rest near the saturation value. We now investigate whether this time delay in N release can explain the observed scatter in N/O, as many have speculated, under conditions of continuous as well as bursting star formation, where the former can be viewed as an infinite number of overlapping bursts. 

\subsubsection{Continuous Star Formation}

We have performed four simulations, designated 1, 2, 3, and 4, respectively, which differ from one another in the range of allowed values for star formation efficiency (SFE) and age, as well as assumptions about instantaneous recycling. Details of these and all other simulations to be discussed are given in Table~1, where the simulation number is listed in the first column, with subsequent columns indicating relevant parameter ranges for each simulation. Results of simulations 1-4 are shown in Figs.~\ref{sim1}-\ref{sim4}. In these plots and others that follow later, abundances derived by Nava et al. are shown with filled circles while plus signs and stars show unsmeared and smeared model points, respectively. A representative error bar is included in the lower right corner of each graph.

Simulation~1 comprises simple models only, i.e. no inflow or 
outflow of matter occurs and instantaneous recycling is assumed, with values of 
SFE and age chosen randomly in each case from within the ranges of 0.01-0.1 
Gyr$^{-1}$ and 0.02-2 Gyr, respectively. The results of this simulation for the N/O-O/H plane are shown in Fig.~\ref{sim1}. The unsmeared points 
associated with simulation~1 roughly fall along a line parallel to the O/H axis within a 
horizontal range which is largely determined by the SFE and age ranges, with the latter 
quantity playing the bigger role. This point is confirmed in Fig.~\ref{o2hvt}, 
which shows a plot of O/H versus age. We see that there is a general 
increase in O/H with age, as expected from the basic notion of chemical 
evolution. The vertical scatter is clearly produced by the additional influence 
of the SFE on buildup of oxygen, since for a given age O/H will be augmented by 
a large SFE, diminished by a small one.

Along the vertical axis of Fig.~\ref{sim1}, unsmeared model points fall at a nearly constant value of -1.35 for simulation~1, a level set by 
the ratio of the integrated yields for these two elements and essentially independent of SFE and age. Thus, the unsmeared points alone fail to reproduce the observations. 

\begin{figure}
\centering
\figurenum{2}
\includegraphics[width=12cm,angle=270]{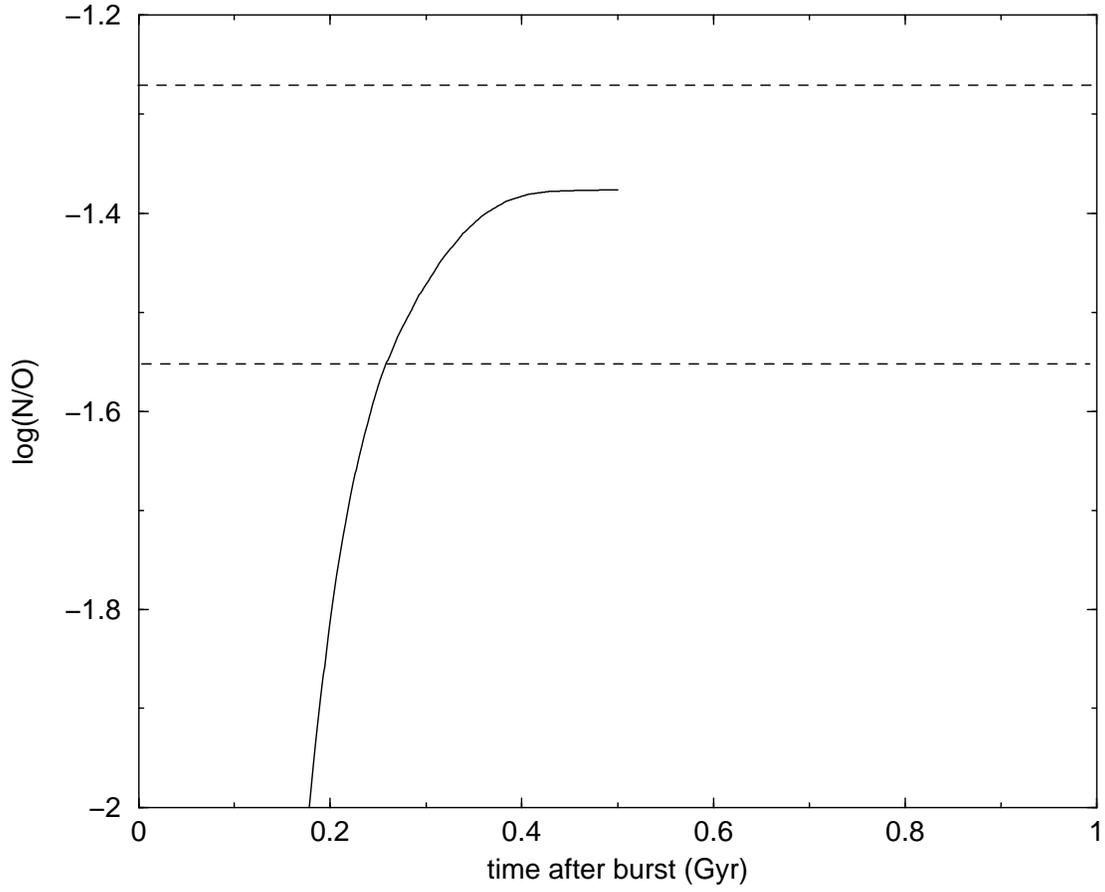}
\caption{Plot of log(N/O) versus the time in Gyr since the beginning of a starburst of duration 100~Myr. The dashed lines show the plateau limits of N/O in Fig.~\ref{aida_final}.}
\label{singleburst}
\end{figure}

\begin{figure}
\centering
\figurenum{3}
\includegraphics[width=12cm,angle=270]{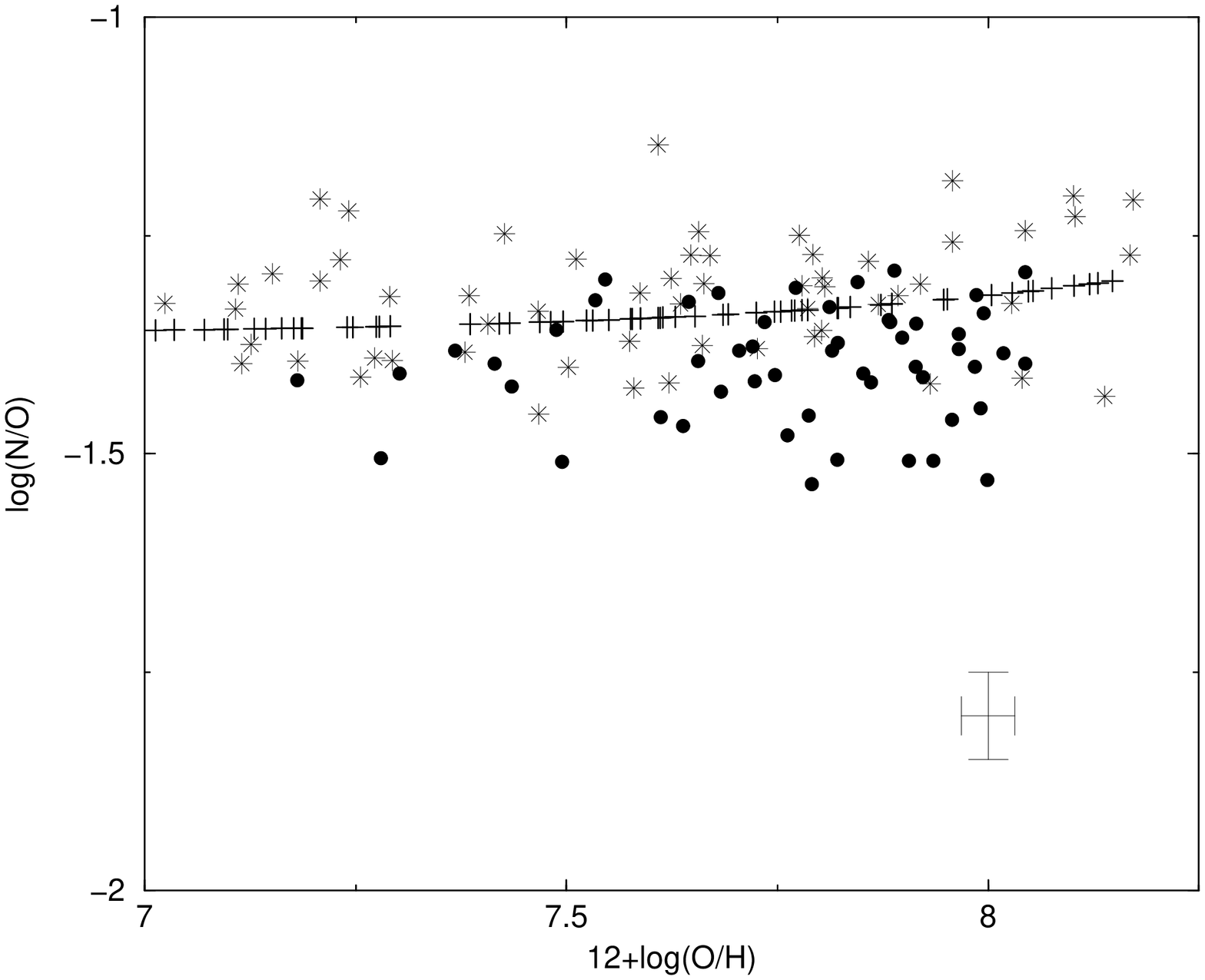}
\caption{Plot of log(N/O) versus 12+log(O/H) for simulation 1. Both unsmeared (plus signs) and smeared (stars) results are shown. 
Sample objects belonging to the N/O plateau as defined by Nava et al. (2006), 
i.e. those objects falling between the horizontal dashed lines in Fig.~\ref{aida_final}, 
are shown as filled circles but with error bars suppressed. A representative set of error bars is shown
in the lower right region of the graph.}
\label{sim1}
\end{figure}

\begin{figure}
\centering
\figurenum{4}
\includegraphics[width=12cm,angle=270]{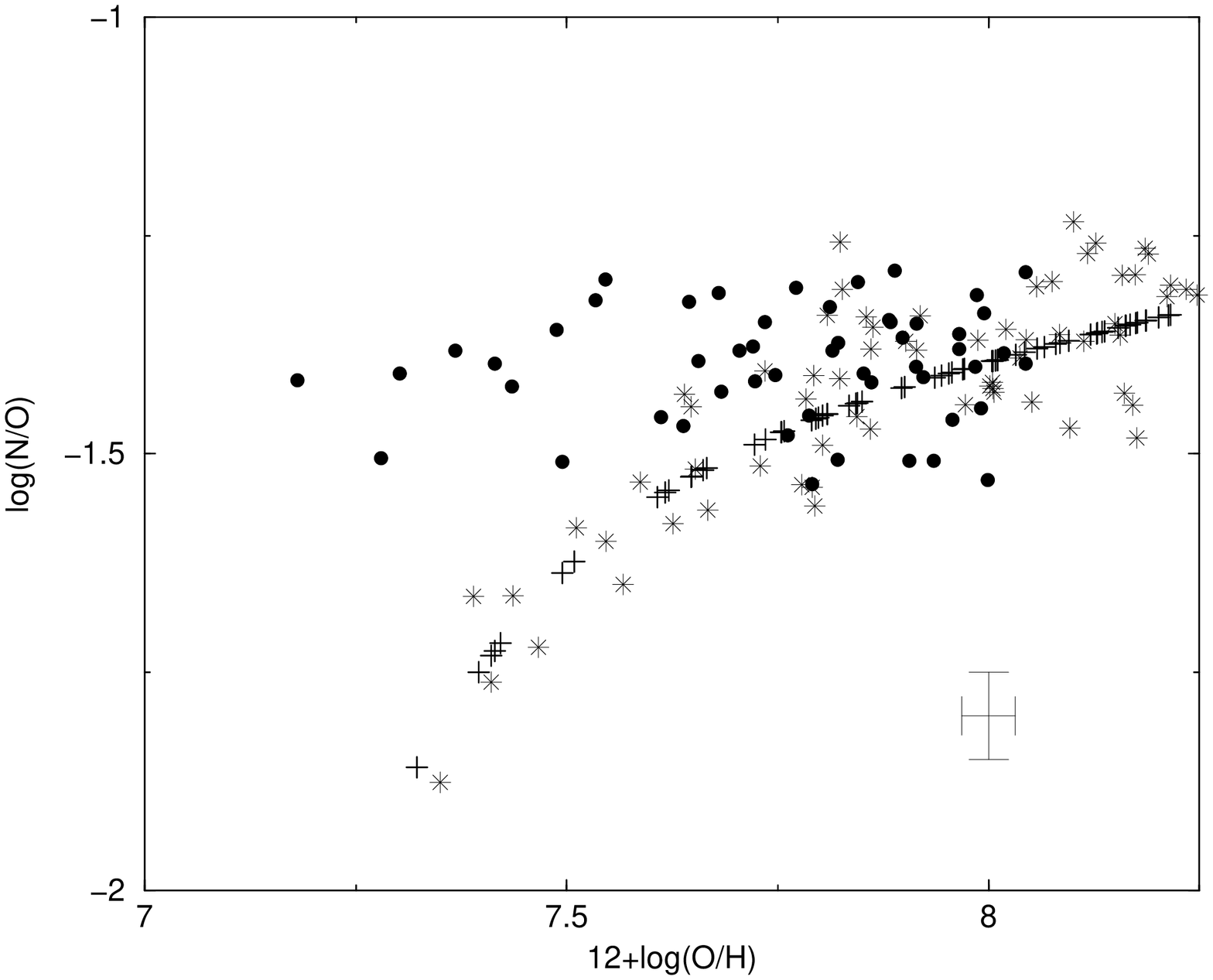}
\caption{Plot of log(N/O) versus 12+log(O/H) for simulation 2. Both unsmeared (plus signs) and smeared (stars) results are shown. 
Sample objects belonging to the N/O plateau as defined by Nava et al. (2006), 
i.e. those objects falling between the horizontal dashed lines in Fig.~\ref{aida_final}, 
are shown as filled circles but with error bars suppressed. A representative set of error bars is shown
in the lower right region of the graph.}
\label{sim2}
\end{figure}

\begin{figure}
\centering
\figurenum{5}
\includegraphics[width=12cm,angle=270]{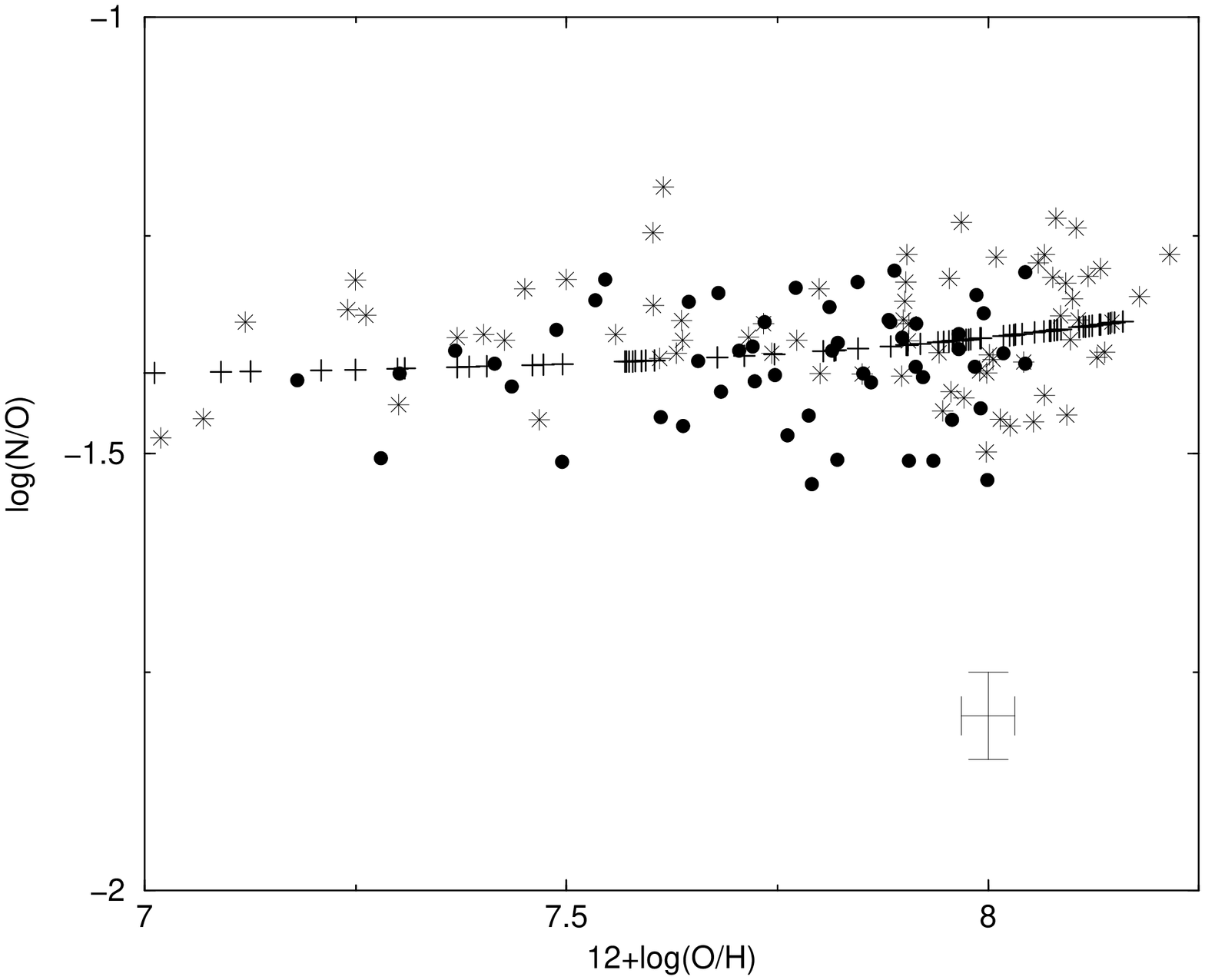}
\caption{Plot of log(N/O) versus 12+log(O/H) for simulation 3. Both unsmeared (plus signs) and smeared (stars) results are shown. 
Sample objects belonging to the N/O plateau as defined by Nava et al. (2006), 
i.e. those objects falling between the horizontal dashed lines in Fig.~\ref{aida_final}, 
are shown as filled circles but with error bars suppressed. A representative set of error bars is shown
in the lower right region of the graph.}
\label{sim3}
\end{figure}

\begin{figure}
\centering
\figurenum{6}
\includegraphics[width=12cm,angle=270]{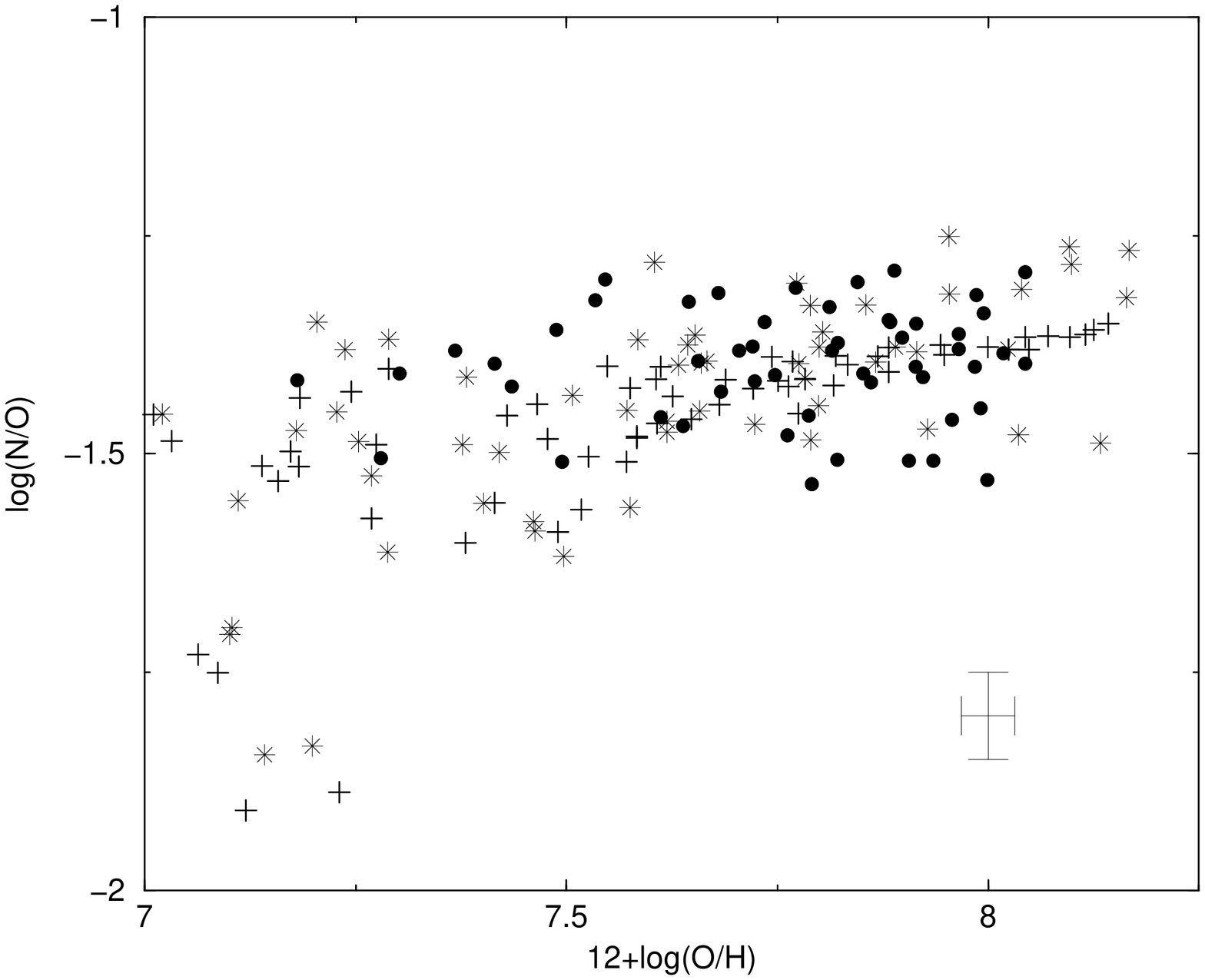}
\caption{Plot of log(N/O) versus 12+log(O/H) for simulation 4. Both unsmeared (plus signs) and smeared (stars) results are shown. 
Sample objects belonging to the N/O plateau as defined by Nava et al. (2006), 
i.e. those objects falling between the horizontal dashed lines in Fig.~\ref{aida_final}, 
are shown as filled circles but with error bars suppressed. A representative set of error bars is shown
in the lower right region of the graph.}
\label{sim4}
\end{figure}

\begin{figure}
\centering
\figurenum{7}
\includegraphics[width=12cm,angle=270]{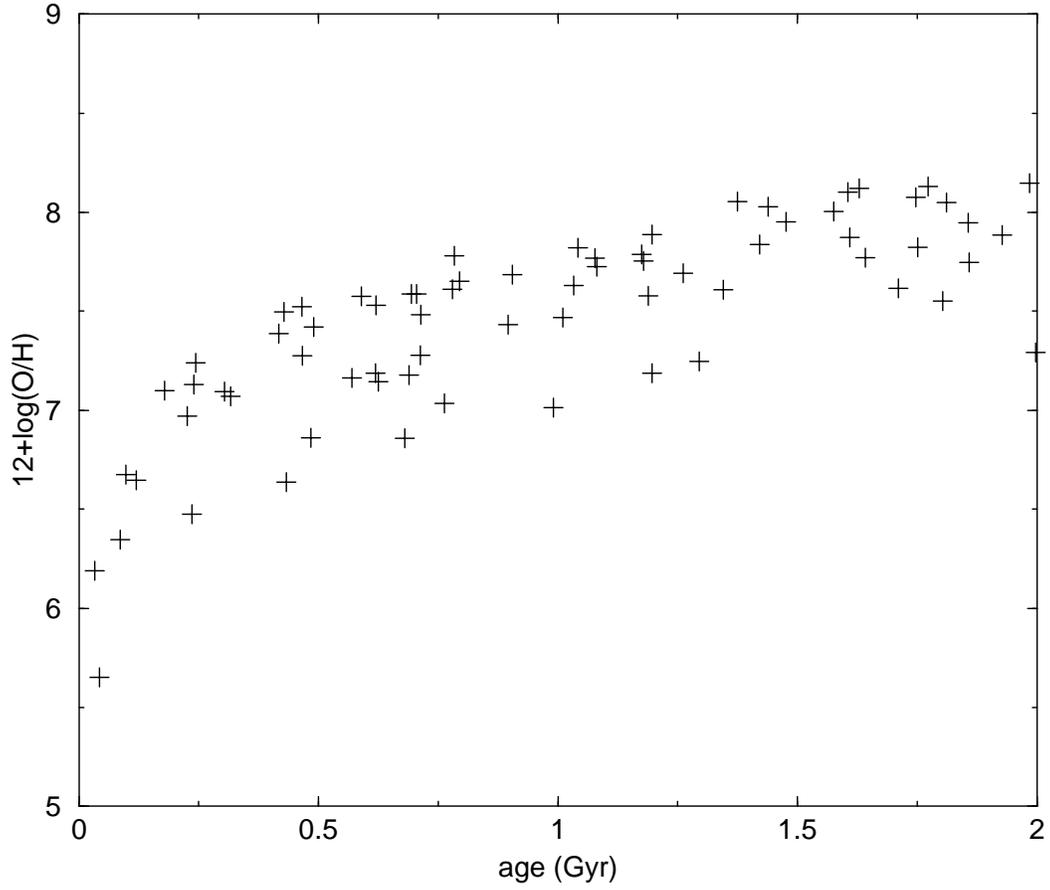}
\caption{Plot of 12+log(O/H) versus galaxy age since the beginning of star formation in Gyr for simulation~1.}
\label{o2hvt}
\end{figure}

For the simulations 2, 3, and 4 we relax the instantaneous recycling approximation in 
order to introduce the effects of time delay on N buildup demonstrated in Fig.~\ref{singleburst}. In the specific case of simulation~2 we hold the SFE constant at 0.1~Gyr$^{-1}$ but allow galaxy age to vary randomly between 0.1 and 2.0~Gyr. The new results are shown in Fig.~\ref{sim2}. With a non-varying SFE, metallicity and age are closely tied. We see for unsmeared simulation~2 models that young, low metallicity objects show clear evidence of delay in N release, as their N/O values are well below the simple model level. On the other hand, systems with higher metallicity are older and N/O is greater, as expected, and eventually the behavior of N/O with O/H flattens as saturation occurs. Note, however, that the saturation level is below the integrated yield ratio of the simple models, because with continuous star formation N can never completely catch up to O. Only when star formation shuts down will that occur. Another result to notice is the distribution of points for simulation~2. Relatively few objects are found at low metallicities because the time required for objects to evolve through this region, going from N/O of -2 to -1.5, say, is roughly 100~Myr (see  Fig.~\ref{singleburst}). Above this value N/O increases more slowly, and consequently objects congregate around the nearly-horizontal portion of the track. However, like simulation~1, simulation~2 fails to produce a good match to the distribution of the observational data, because galaxies continue to fall along a narrow track.

Next, we performed a simulation in which galaxy age is held constant at 2~Gyr while the SFE varies between 0.005 and 0.09~Gyr$^{-1}$. Results for simulation~3 are shown in Fig.~\ref{sim3}. Since we are sampling objects at an age which exceeds the N/O transition time shown in Fig.~\ref{singleburst}, as expected the distribution of N/O for unsmeared models is flat; the SFE only affects progress along the horizontal axis, and so simply varying the SFE by itself does not explain the vertical scatter.

Finally, simulation~4 (Fig.~\ref{sim4}) shows the results when {\it both} SFE and age are selected at random and instanteous recycling remains relaxed. We see that much more scatter is now present at low metallicities, since we have essentially taken the distribution of simulation~2 and spread it out in the horizontal direction by randomly choosing the SFEs. At higher O/H and N/O levels, however, the scatter is less, and this result arises for two reasons. First, objects at higher metallicities are usually older, and hence the gap between N and O expulsion that is caused by delay has narrowed considerably, so the simulation does not predict objects with both low N/O values and advanced ages. In addition, as N/O rises toward the simple model value, vertical progress slows, as can be seen in Fig.~\ref{singleburst}, where the curve begins to turn over as the slope decreases with time. This is entirely expected, since the proportion of new N being expelled from stars is much smaller at large N, although the absolute amount remains roughly constant. So at higher metallicity, simulation~4 predicts no objects with low N/O; instead the points are bunched up near the saturation level of simulation~3.

A potential solution to this problem occurring at high metallicity might be to assume that all 
systems are younger than 500~Myr in order to take advantage of delay while 
increasing the median SFE to obtain the same metallicity spread. However, such an assumption is contrary to observational evidence which indicates that these systems are at least 1~Gyr old; indeed some estimates place them closer to 
10~Gyrs (Skillman et al. 2003). {\it We conclude that when metallicities exceed 7.8 and star formation is continuous, delay in N release by IMS is an unlikely explanation for the observed scatter in 
N/O.} This result conflicts with the conclusions of HEK but is in agreement with the findings of 
Lanfranchi \& Matteucci (2003).

From results of the first four experiments involving continuous star formation, it is clear that if observational uncertainty is ignored: 1)~an object's position along the 
horizontal axis of the N/O-O/H plot is determined by the product of the SFE and 
system age, as expected; and 2)~an object's position along the vertical axis is controlled by 
the ratio of the integrated yields and time delay of N release.

\subsubsection{Bursting Star Formation}

The unsmeared models of simulations 1-4 above show that the effects of delay can only produce spread in N/O if 
a significant number of objects in the observed sample have not advanced in age 
by more than about 500~Myr since the current episode of star formation began. In 
the case of continuous star formation, this occurs only once in a system's 
lifetime when it is very young. However, if a system's star 
formation history is characterized by bursting episodes, then this 
increases the likelihood that objects will be observed close enough in time to a 
burst that effects of delay will spread the sample out in N/O.

\begin{figure}
\centering
\figurenum{8}
\includegraphics[width=12cm,angle=270]{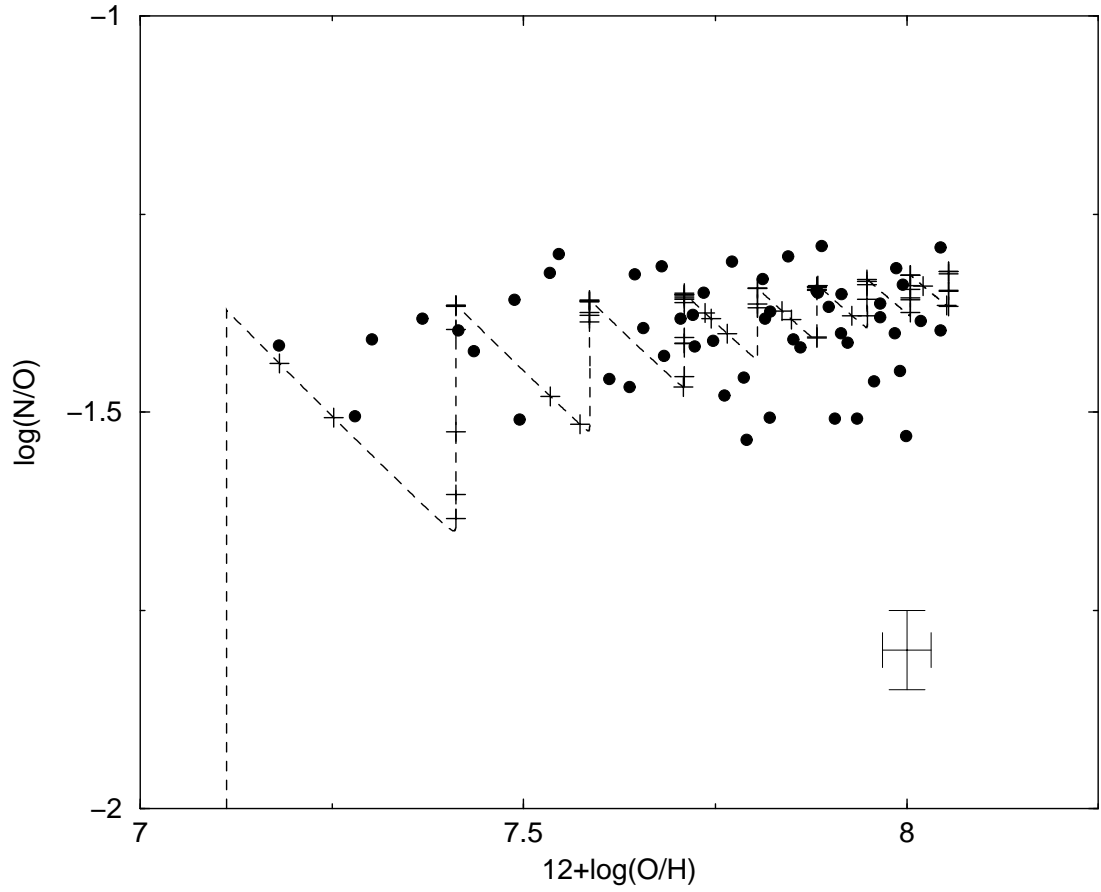}
\caption{Plot of log(N/O) versus 12+log(O/H) for the multiple burst models of simulation~5. The dashed line indicates the complete
evolutionary path followed by an object over time as bursts erupt. The filled circles show the observations.}
\label{sim5line}
\end{figure}

To illustrate this with a simple example, we produced simulation~5, 
where 
for each object in the sample  a new burst began every 500~Myr and lasted 100~Myr. Object
ages were chosen randomly from a range of 1-5~Gyr. In each case the SFE was held
at a constant 0.075~Gyr$^{-1}$. The result is shown in Fig.~\ref{sim5line}. 

The sawtooth 
line shows the complete model track that would be followed by a single system as it
evolved over 5~Gyr. Simulation~5 simply samples objects at different times along
this line.
As a burst occurs, oxygen (but not nitrogen) is immediately expelled, and
the line moves down and 
to the right. When the delayed N is released, the track moves upward vertically until the release is
complete, at which time progress is halted until the next burst. Since object ages were
selected at random, more points will be found along track segments 
corresponding to slow evolution. Thus, a majority of points tend to collect at 
the top of the vertical segments where the system is temporarily dormant and waiting for the 
next burst to occur, while a much smaller number populate the diagonal and lower extremes of 
the track; evolution here is progressing rapidly.

Clearly, at low metallicities, bursting may potentially explain the N/O scatter.
However, the predicted clustering of points at higher 
metallicities, where objects 
tend to congregate at the higher N/O values, as discussed earlier, is not 
seen in the observations. But a more serious problem with bursting is that as
the absolute abundances of N and O increase, the vertical excursions become
relatively smaller. This apparent dampening is, of
course, the bursting analog of the effect seen for simulation~4 in Fig.~\ref{sim4},
where a constant amount of newly ejected N adds to an increasing level of this element.

\begin{figure}
\centering
\figurenum{9}
\includegraphics[width=12cm,angle=270]{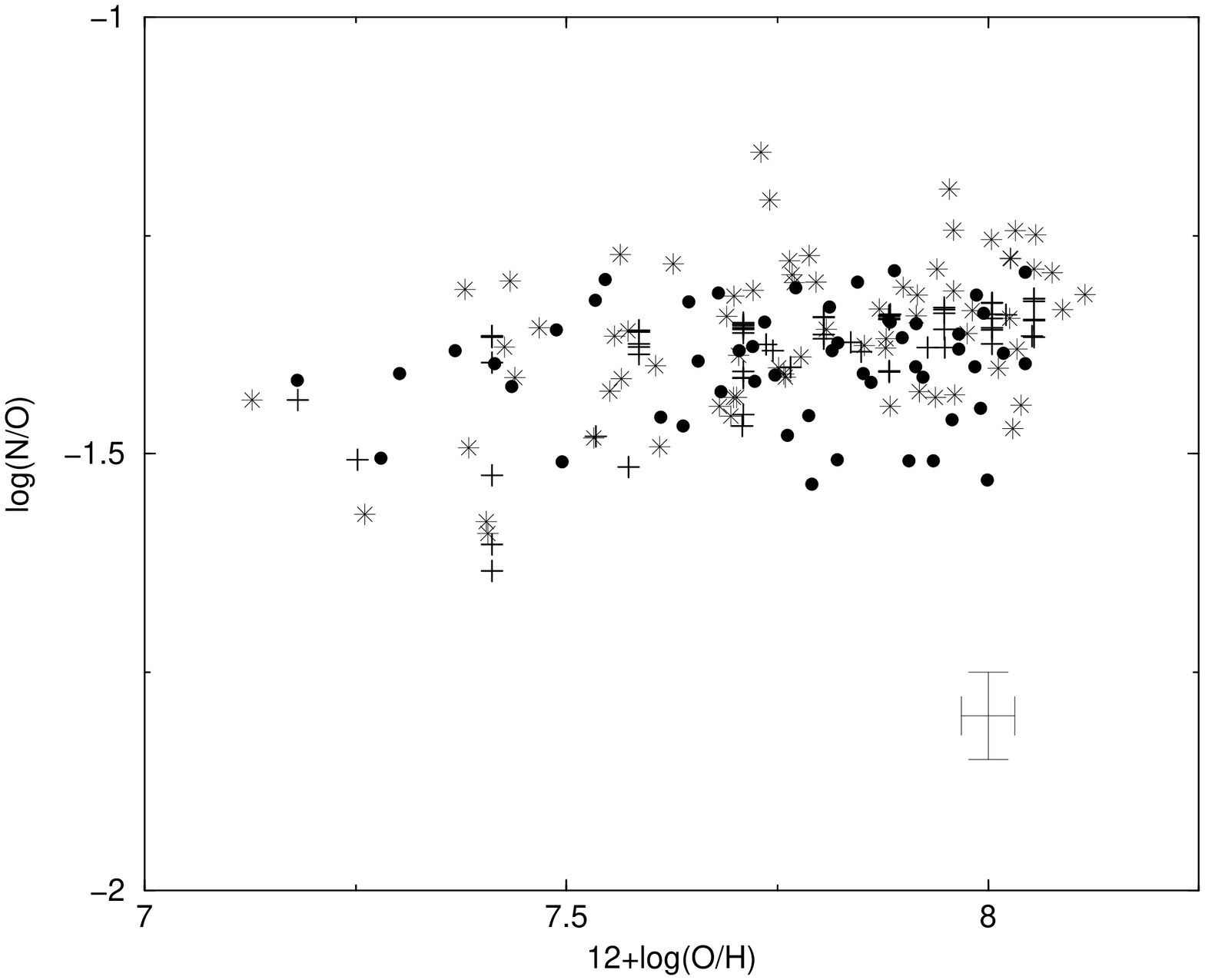}
\caption{Plot of log(N/O) versus 12+log(O/H) for simulation 5. Both unsmeared (plus signs) and smeared (stars) results are shown. 
Sample objects belonging to the N/O plateau as defined by Nava et al. (2006), 
i.e. those objects falling between the horizontal dashed lines in Fig.~\ref{aida_final}, 
are shown as filled circles but with error bars suppressed. A representative set of error bars is shown
in the lower right region of the graph.}
\label{sim5}
\end{figure}

\begin{figure}
\centering
\figurenum{10}
\includegraphics[width=12cm,angle=270]{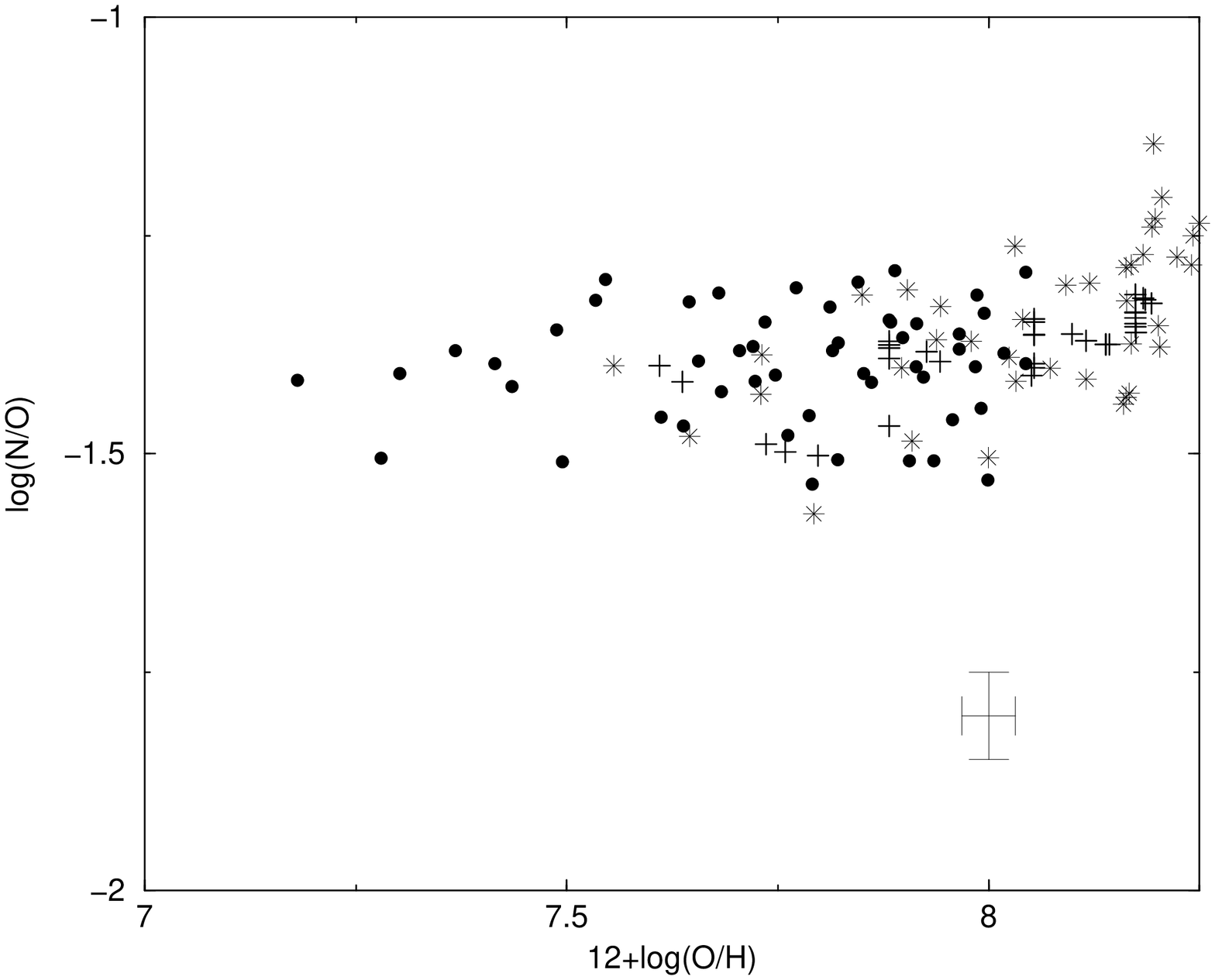}
\caption{Plot of log(N/O) versus 12+log(O/H) for simulation 6. Both unsmeared (plus signs) and smeared (stars) results are shown. 
Sample objects belonging to the N/O plateau as defined by Nava et al. (2006), 
i.e. those objects falling between the horizontal dashed lines in Fig.~\ref{aida_final}, 
are shown as filled circles but with error bars suppressed. A representative set of error bars is shown
in the lower right region of the graph.}
\label{sim6}
\end{figure}

\begin{figure}
\centering
\figurenum{11}
\includegraphics[width=12cm,angle=270]{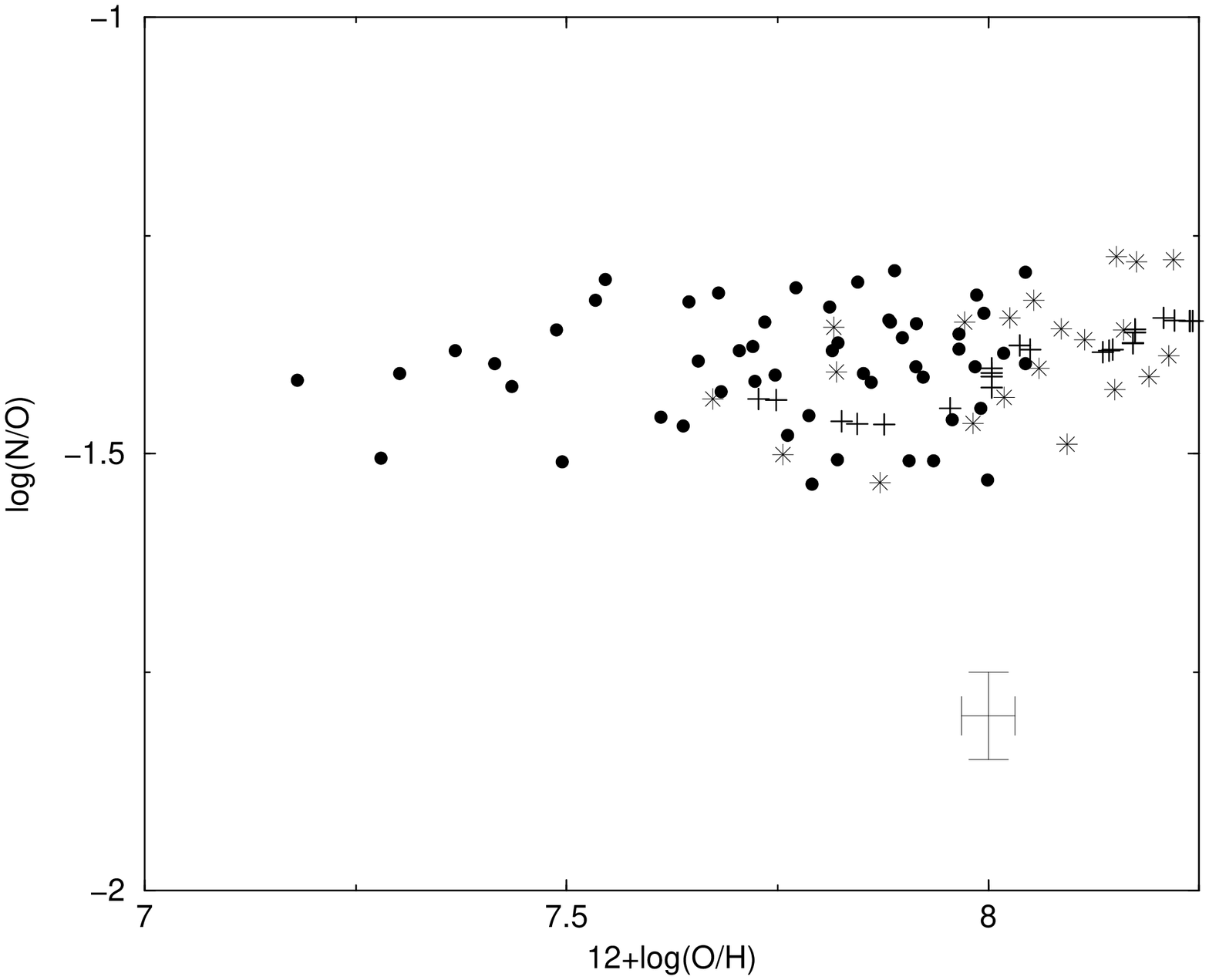}
\caption{Plot of log(N/O) versus 12+log(O/H) for simulation 7. Both unsmeared (plus signs) and smeared (stars) results are shown. 
Sample objects belonging to the N/O plateau as defined by Nava et al. (2006), 
i.e. those objects falling between the horizontal dashed lines in Fig.~\ref{aida_final}, 
are shown as filled circles but with error bars suppressed. A representative set of error bars is shown
in the lower right region of the graph.}
\label{sim7}
\end{figure}

The damping problem might be overcome if the burst {\it duration} is allowed to be longer. This would cause
objects to attain higher metallicities earlier than before and effectively fill the void that results from the small range in N/O and metallicities exceeding about 7.70. To explore this option, we calculated simulations~6 and 7, in which the burst
durations were set at 300~Myr and 400~Myr, respectively. In Fig.~\ref{sim5}-\ref{sim7} we show
the results of simulations~5, 6, and 7 in order. We note that as burst duration is lengthened, the vertical spread is ostensibly improved, but only because the entire group of points is progressively shifted to higher metallicity. That is, increasing the
burst duration essentially stretches the sawtooth in Fig.~\ref{sim5line} horizontally. However, there is now a paucity of points at low metallicity. So far, then,
the observations cannot be reproduced at high metallicities. 

We explored other possibilities for understanding the morphology of the N/O plateau in terms of bursting episodes by
running many additional simulations.
In the end, we found that varying burst duration or frequency, SFE, or 
age separately or together failed to reproduce the desired scatter, because all of 
these changes simply slowed or accelerated the rate of chemical evolution but 
could not overcome the damping problem.

In the absence of an obvious combination of parameters which are capable of producing
a sample of objects which matches the observed distribution of objects in the
N/O-O/H plane, {\it we conclude that when observational scatter is ignored, bursting alone is unlikely to explain plateau morphology, at least at O/H values exceeding 7.8}. Together with our earlier results for continuous star formation,
it seems unlikely that the full explanation for the vertical distribution of
galaxies involves star formation history exclusively. This conclusion differs somewhat with the findings of Chiappini, Romano, \& Matteucci (2003) concerning BCGs. These authors calculated bursting models using different burst onset times and durations as well as different star formation histories and concluded that time delay can explain the scatter above O/H of 7.8. In particular, we are unable to produce any models with N/O values above -1.2, such as their models E, K, and F\footnote{Indeed this is one of the reasons we elected to exclude the high N/O objects at O/H$>$7.9 from consideration. Not only did they seem to fall outside of an otherwise Gaussian distribution and well above the narrow plateau distribution, but they seemed to require an additional mechanism beyond what is necessary for the other plateau objects to explain their high N/O values.}; this point may be related to their use of the v.d. Hoek \& Groenewegen (1997) yields for IMS. On the other hand, we agree with the conclusion of Chiappini et al. (2003) that primary N production by massive stars is unnecessary for explaining the N/O plateau objects when observational scatter is ignore (see \S3.3 below).

\subsection{Effective Yields and Outflow}

Another general type of explanation which could potentially explain the morphology of the N/O-O/H plane is a variation in the effective yield of alpha elements. In this section we will test a specific aspect of this variation, that related to the selective expulsion of these elements, including oxygen, from sample galaxies by galactic winds whose intensities vary from galaxy to galaxy. Element loss in this manner results in a reduction of 
their effective yields, and thus by allowing the amount of loss to vary among objects, the sample of objects may be expected to exhibit a range of N/O ratios values. 
Varying the effective yield in dwarf irregulars was proposed early-on by Matteucci \& Chiosi (1983), Matteucci \& Tosi (1985), 
and Matteucci (1986b) 
as a mechanism for explaining the observed abundance patterns. More recently, chemical evolution models including winds have been successfully employed by Marconi, Matteucci, \& Tosi (1994) and Romano, Tosi, \& Matteucci (2005) to explain the scatter in N/O for BCGs.

Selective element loss is assumed to be linked to Type~II supernova events, when stars in excess of 8~M$_{\odot}$ explode. We include this in our calculations simply by multiplying the massive star contribution to the right hand side of Eq.~\ref{ex} by (1-wl), where wl is the windloss parameter chosen at random in our simulations from a specified range. In addition, appropriate adjustments were made to Eq.~\ref{g} to account for the bulk removal of gas from a galaxy.

Simulation 8 is the result of randomly selecting windloss parameters from the range between 0 and 0.5, a range which we
determined by trial and error to be appropriate for producing the observed scatter in N/O. In this simulation,
all objects were considered to be 1~Gyr in age and SFEs were randomly chosen from within the range 
of 0.02-0.2~Gyr$^{-1}$ to reproduce the observed range in O/H. The important properties of simulation~8 are summarized in Table~1\footnote{From the results already described, 
we know that adjustments to age and SFE affect O/H but have little impact on the N/O ratio. 
Therefore, we set the age at a constant 1~Gyr and then experimented with various ranges in SFE until we found one which reproduced the horizontal scatter.}.

\begin{figure}
\centering
\figurenum{12}
\includegraphics[width=12cm,angle=270]{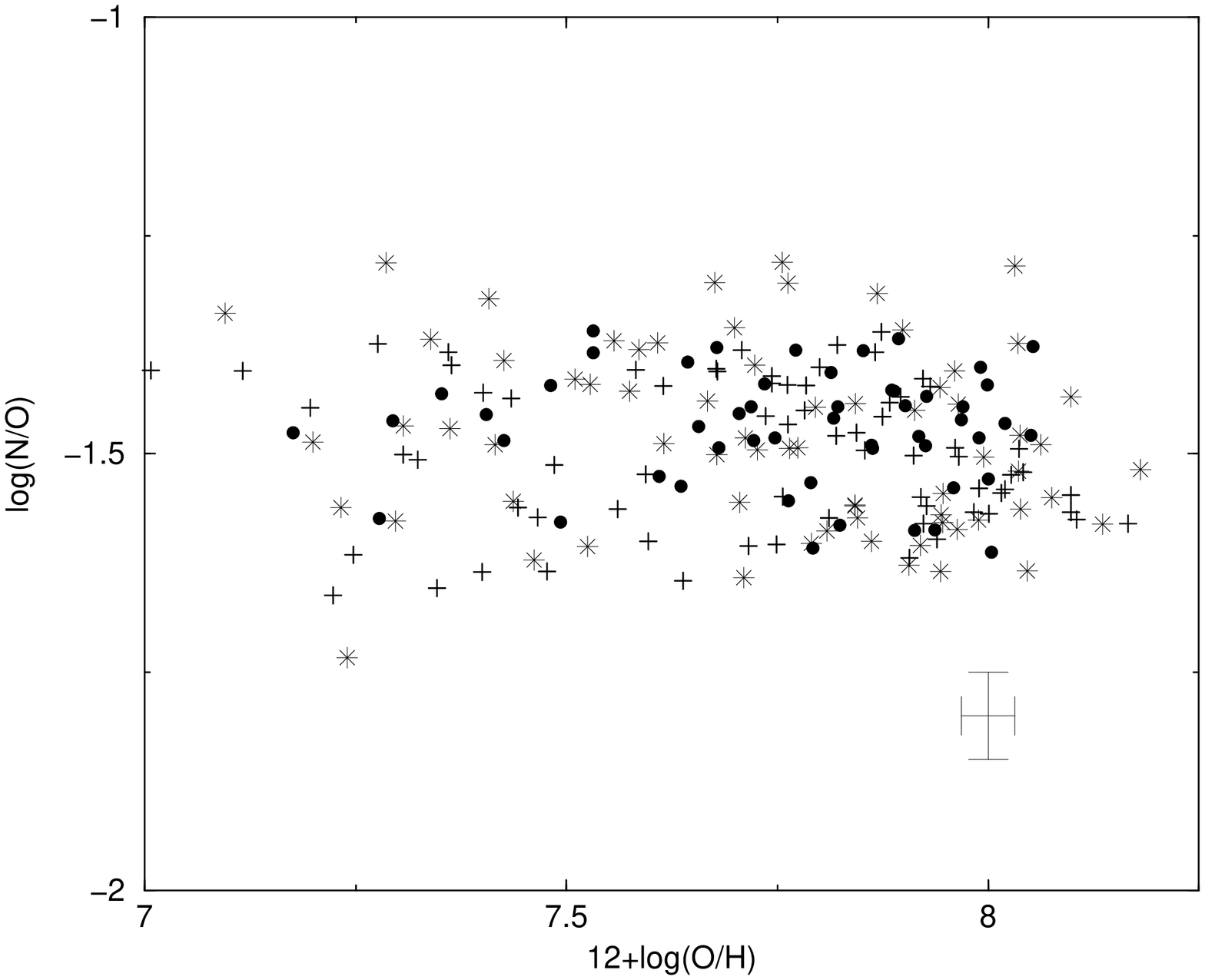}
\caption{Plot of log(N/O) versus 12+log(O/H) for simulation 8. Both unsmeared (plus signs) and smeared (stars) results are shown. 
Sample objects belonging to the N/O plateau as defined by Nava et al. (2006), 
i.e. those objects falling between the horizontal dashed lines in Fig.~\ref{aida_final}, 
are shown as filled circles but with error bars suppressed. A representative set of error bars is shown
in the lower right region of the graph.}
\label{sim8}
\end{figure}

The results for simulation 8 are shown in Fig.~\ref{sim8}. The major result here is that the
vertical spread for the unsmeared models is relatively uniform across the metallicity range under consideration and can be controlled effectively by adjusting the range in the 
windloss parameter. In addition, it was necessary to scale the N yields from IMS by 0.70 in order to center the system of simulated points on the observed data\footnote{While this adjustment may appear to be contrived,
we point out that model-predicted stellar yields are generally taken to be good only to within a factor of 2.
In any case, we are simply testing the feasibility of the outflow mechanism for reproducing the spread in
N/O, so the systematic vertical shift is not cause for concern.}. Finally, we should point out that the
horizontal spread of the simulated points is controlled by the range in SFE, since all objects have the
same age. However, since outflow affects the O abundance, the ranges in SFE and outflow are linked to some
extent, and so in the end a set of compatible ranges in these two parameters could only be found through iterative procedures, i.e. alternately adjusting the SFE and windloss parameters until a suitable match to the data resulted. Thus, our models seem to rule out a large portion of parameter space for these two parameters.

\begin{figure}
\centering
\figurenum{13}
\includegraphics[width=12cm,angle=270]{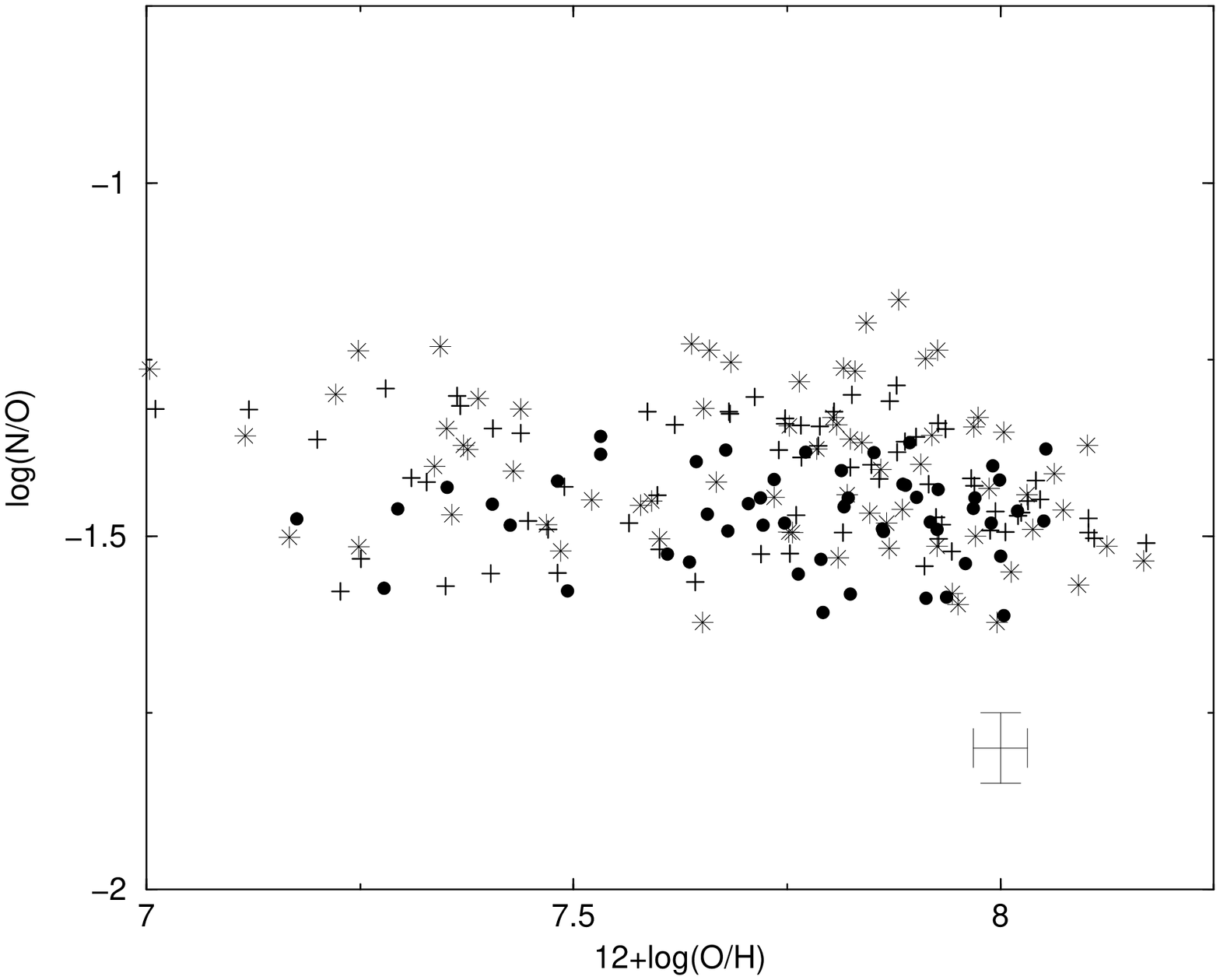}
\caption{Plot of log(N/O) versus 12+log(O/H) for simulation 9. Both unsmeared (plus signs) and smeared (stars) results are shown. 
Sample objects belonging to the N/O plateau as defined by Nava et al. (2006), 
i.e. those objects falling between the horizontal dashed lines in Fig.~\ref{aida_final}, 
are shown as filled circles but with error bars suppressed. A representative set of error bars is shown
in the lower right region of the graph.}
\label{sim9}
\end{figure}

In order to gauge the sensitivity of the above result on age and SFE range, we ran a second simulation, 
simulation~9, again using
outflow, but this time setting all object ages at 10~Gyr, which necessitated a downward adjustment in the SFE 
range to 0.002-0.02. Here, the IMS nitrogen yield was scaled by 0.60.
The results for simulation~9 are shown in Fig.~\ref{sim9}. Here we see a distribution in the unsmeared points
which is remarkably similar to the one predicted by simulation~8, and once again the spread is relatively uniform
across the metallicity range. {\it This result simply underscores the independence
of the parameters associated with the star formation history, on the one hand,
from those related to the spread in N/O on the other.}

\begin{figure}
\centering
\figurenum{14}
\includegraphics[width=12cm,angle=270]{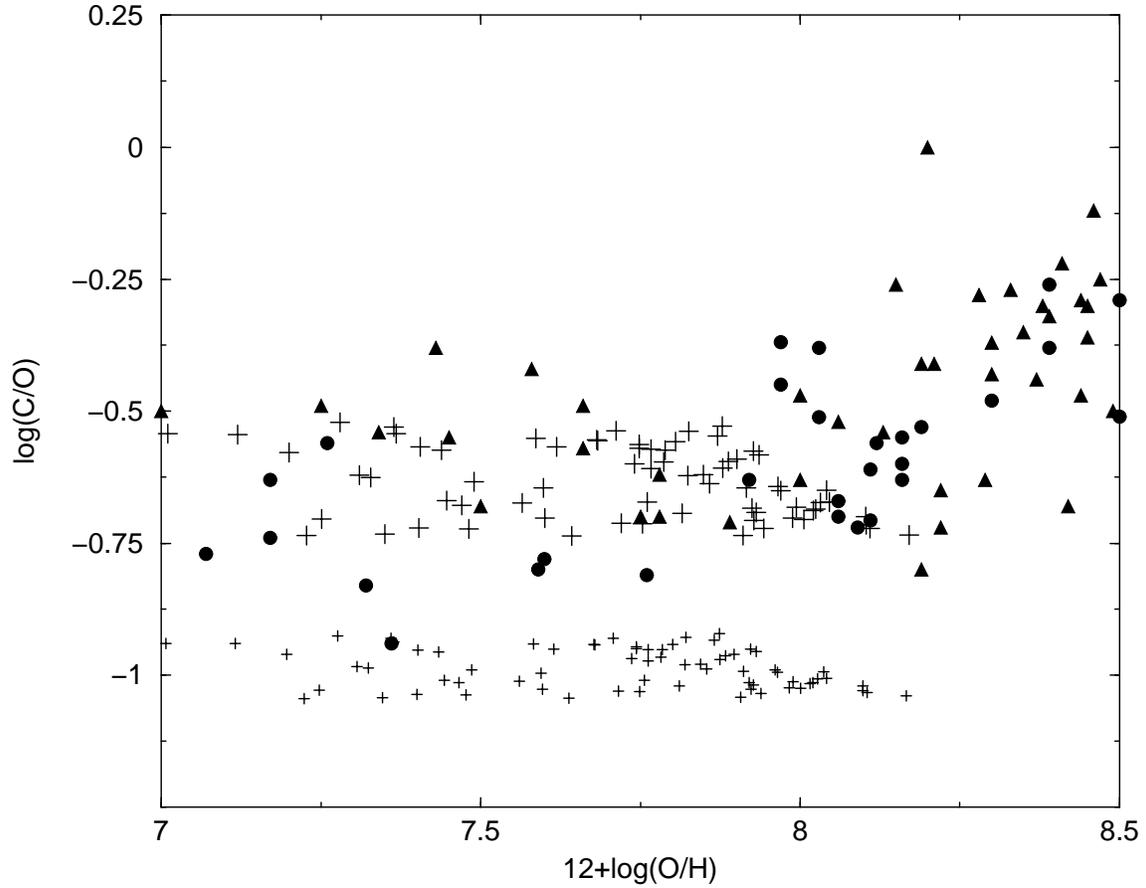}
\caption{Plot of log(C/O) versus 12+log(O/H) for simulations 8 (small plus signs) and 9 (large plus signs). 
Observations of extragalactic emission line objects (filled circles) are taken from Garnett et al. (1995), Izotov \& Thuan (1999), 
and Kobulnicky \& Skillman (1996). Stellar data (filled triangles) are taken from Gustafsson et al. (1999) and 
Gummersbach et al. (1998).}
\label{c2o}
\end{figure}

As outflow appears to be more promising than delay for producing a uniform point distribution, we extended our outflow analysis to other element ratios for simulations~8 and 9. Fig.~\ref{c2o} is a plot of 
C/O versus O/H, showing predictions for the C/O ratio for these two simulations as well as observational 
results for emission line objects (filled circles) from Garnett et al. (1995), Izotov \& Thuan (1999), and Kobulnicky \& Skillman (1996) and stellar data (filled triangles) from Gustafson et al. (1999; solar-type MW disk stars) and Gummersbach et al. (1998; B stars in the MW disk). Both IMS and massive star C yields were scaled by 0.30 in these simulations in order to match the data better.
There is a clear difference in
predictions of the two simulations. Simulation~9, where object ages were all 10~Gyr, matches the
observations quite nicely, while simulation~8, where the ages were all 1~Gyr, predicts C/O values which are systematically shifted downward by about 0.5~dex
relative to the simulation~9 results.

The vertical offset between simulations~8 and 9 in Fig.~\ref{c2o} is easily explained by the delay in C production which arises because a significant portion of $^{12}$C is produced in stars whose main sequence lifetimes exceed 1~Gyr, i.e. M$\le$2.5~M$_{\odot}$, according to the results of Marigo's yields (see Fig.~1.2 and Table~1.2 in Henry 2004). This result would seem to rule out young BCGs and favor the scenario implied by the recent work of van~Zee \& Haynes (2006) and by Skillman et al. (2003) which suggests that these objects are really chemically advanced systems of several Gyr in age.

\begin{figure}
\centering
\figurenum{15}
\includegraphics[width=12cm,angle=270]{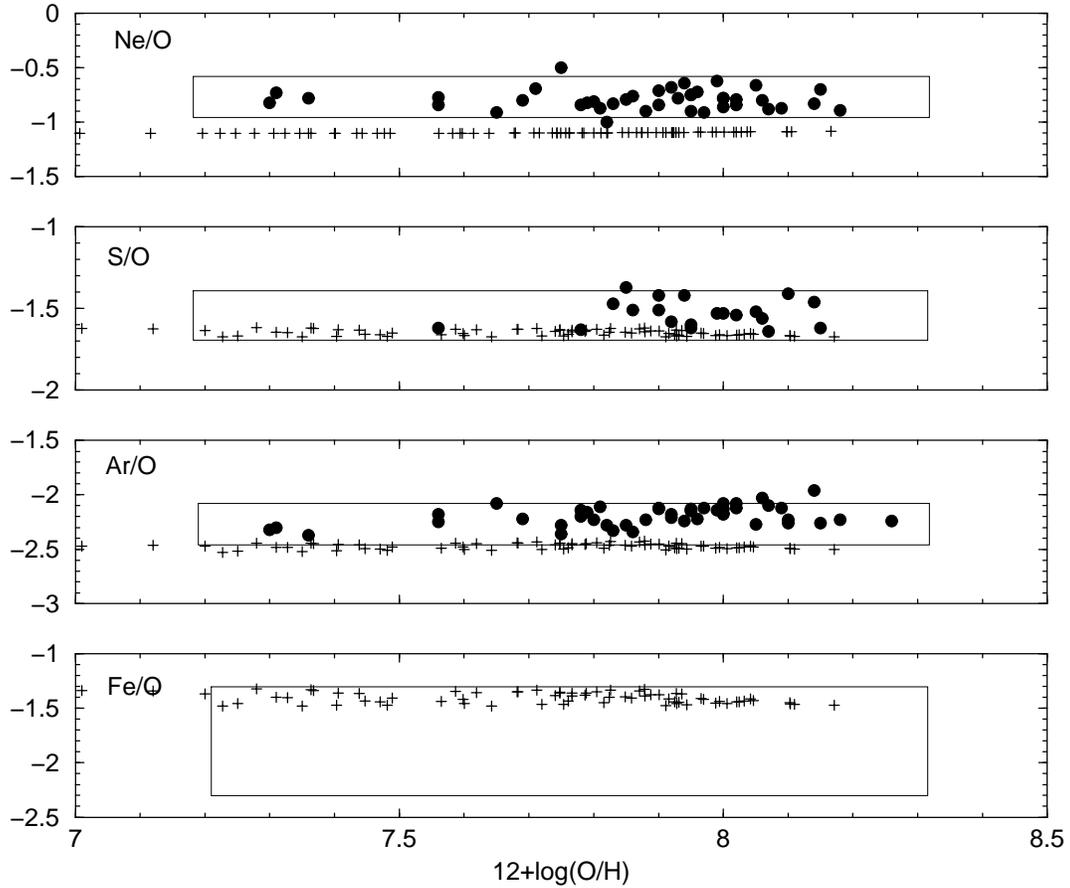}
\caption{Plots of log(Ne/O), log(S/O), log(Ar/O), and log(Fe/O) versus 12+log(O/H) for simulation 8. Observational 
data are from van~Zee and Haynes (2006; filled circles) and the survey by Izotov et al. (2004), whose limits are
indicated with a rectangle in each panel.}
\label{nesarfevo}
\end{figure}

In Fig.~\ref{nesarfevo} we show the predictions of simulation~8 for Ne/O, S/O, Ar/O, and Fe/O\footnote{Since results for simulations 8 and 9 are nearly identical, we only show points for simulation 8.}. (Results for simulation~8 and 9 are essentially identical, so only simulation~8 is shown in the figure.) For comparison we include
the data for isolated dIs from van~Zee \& Haynes (2006; filled circles). The large rectangular box in each
panel shows the limits of the observed values for a large sample compiled by Izotov et al. (2004) comprising dIs taken from the Sloan Digital Sky Survey and BCG results already available. In all but the bottom panel
our simulated points show a tendency to fall below the observed level, perhaps indicating that the O yields of
Portinari et al. (1998) are systematically too large\footnote{Since the top three panels all concern alpha elements, which are expelled by galactic winds in equal proportions, the offset cannot be explained by outflow.}. However, because of uncertainties in stellar yield predictions, we believe that the simulation results are compatible with the observations. The
narrow vertical distribution of points is also expected, since alpha elements are all
presumably forged in the same stellar sites. Likewise, the good agreement between theory and observation
for Fe/O is not surprising, since galaxy ages of 1~Gyr (simulation~8) or 10~Gyr 
exceeds the delay in Fe release by Type~Ia supernovae.

In summary, for models as yet unsmeared for observational uncertainties, the observed vertical spread in N/O for BCGs appears consistent with a range in outflow rates for alpha elements. The use of outflow, however, may be inconsistent with the observed direct relation between luminosity and
metallicity for dIs (van~Zee et al. 2006; Lee et al. 2003; 
Tremonti et al. 2004). If luminosity is assumed to trace galaxy mass, then objects of greater metallicity
are generally more massive, and outflow might be less likely to occur in these cases due to their 
greater gravitational potentials. The expectation, then, is diminished vertical spread for the more metal-rich
dIs. In fact, just the opposite may be the case (see Fig.~\ref{aida_final}). While this may pose a serious challenge to the outflow mechanism for explaining the plateau morphology, we point
out that the interpretation and implications of the luminosity-metallicity relation with respect to chemical evolution models are still controversial.
In addition, outflow may be controlled more by
the {\it local} surface potential than by the global one, and these two potentials may differ greatly from one another. Thus, it is too early to exclude outflow from consideration based only upon the luminosity-metallicity relation.

Another process related to the variation of the effective yield and which
may influence the time evolution of metallicity from object to object in a way similar to outflow is the efficiency with which newly ejected materials mix with the surrounding interstellar medium. In fact in our models we have assumed that homogeneity occurs instantaneously, although this may not always be the case. Kobulnicky \& Skillman (1996) studied numerous H~II regions surrounding WR stars and failed to detect regions of local enrichment which would indicate the lack of complete mixing of material as it is expelled from stars. On the other hand, the time scale for mixing may exceed a few million years due to barriers to matter flow set up by either global or local magnetic fields. Despite the large amount of work which has gone into this subject (Edmunds 1975; Roy \& Kunth 1995; Elmegreen 1998; Daflon et al. 1999; Oey 2000; de~Avillez \& Mac~Low 2002), the actual effects of 
mixing remain unclear, and so we don't consider this mechanism further.

Finally, throughout our analysis we have purposely excluded from consideration the small number of points at relatively high metallicity ($>$7.8) in Fig.~\ref{aida_final} which fall above what we define as the N/O plateau. We did this for two reasons. First, the distribution of N/O values in Fig.~8 of Nava et al. (2006), indicates that these points form a separate, less populated group, while the objects on the plateau as we define it are larger in number and seem to form a Gaussian distribution of their own. Thus, there is a suggestion that the two groups may be the product of different star formation histories or effective yield schemes. Second, the results of our analysis in this paper indicates that in order to accomodate these objects in our simulations, an additional process would have to switch on at metallicities above a threshold of 7.8. Indeed, we experimented with a few individual chemical evolution models to see what would be required to raise the N/O value enough to explain the high N/O objects. In the end we were able to force model tracks to move into this region by increasing the value of the windloss parameter when the metallicity rose above 7.8. This solution has an {\it ad hoc} quality to it, and thus we did not continue with the analysis. On the other hand, models presented by Chiappini, Matteucci, \& Ballero (2005) and Romano, Tosi, \& Matteucci (2005) which employ IMS yields of van~den~Hoek \& Groenewegen (1997) were successful at predicting these higher N/O values. In addition, a model by Matteucci (1986) in which massive stars were assumed to contribute to primary N production, were also successful. Thus, the existence of these high N/O objects may suggest that higher N yields are more appropriate.

\subsection{The Effects of Observational Scatter}

Despite the fact that Nava et al. have shown that nearly all of the scatter on the N/O plateau is due to observational uncertainty, our entire discussion thus far has centered on the results of Monte Carlo simulations in which potential effects of observational scatter have been ignored. Our purpose has been to offer clear comparisons of the impact that ranges in star formation efficiency, galaxy age, burst frequency and duration, and outflow have on the predicted abundance characteristics of a sample of galaxies in the absence of scatter.

It is now time to introduce observational scatter into the simulations and see if our conclusions change in any way. We have already described in {\S}~2 how the model points within a simulation were smeared to simulate the effects of observational uncertainty. The resulting points are shown with star symbols in the same plots for individual simulations already discussed. We now compare the new results with the unsmeared models.

We begin with simulations 1-4 shown in Figs.~\ref{sim1}-\ref{sim4}. Recall that these simulations refer to situations of continuous star formation. Fig.~\ref{sim1} refers to the simple, closed box scenario. Whereas the unsmeared models display almost no variation in N/O, the introduction of scatter changes the result entirely. The simple model simulation now appears to be a viable candidate for explaining the scatter on the N/O plateau, assuming that the systematically high N/O values could be accounted for by adjusting N yields downward. 
Since the simple models ignore stellar lifetimes, they closely approximate the situation in which rapidly evolving massive stars are the sole producers of nitrogen. The fact that the point spread now resembles that of the observations 
has important implications about the debate over the stellar mass range most responsible for nitrogen production. Recall that Izotov \& Thuan (1999) concluded that massive stars are the principal nitrogen producers. The smeared results of simulation~1 support their contention. 

In Figure~\ref{sim2} when scatter is added to the simulation results we see that the lack of points sufficiently high in N/O to match the data still exists at low metallicity. We conclude from this that simply a range in galaxy age cannot explain the observations. However, systems having the same age but possessing a range of SFEs do seem consistent with the observations, as we see for simulation~3 in Fig.~\ref{sim3}, all because points on the original narrow track are smeared sufficiently, as we also saw in the case of the simple models in Fig.~\ref{sim1}. Likewise, allowing for a range both in SFE and galaxy age simultaneously seems compatible with the data, as seen for simulation~4 in Fig.~\ref{sim4}. So collectively the continuous star formation simulations with scatter included are consistent with object positions on the N/O plateau as we have defined it, with the exception of the case in simulation~2 where only age is allowed to vary. This particular result indicates that if IMS contribute significantly to N production, then galaxy ages exceed 250~Myr, i.e. the time necessary for IMS to eject their nuclear products, since (barring selection effects) no objects are observed below the plateau.

Simulations 5, 6, and 7 concern bursting star formation histories, with the results shown in Figs.~\ref{sim5}-\ref{sim7}. Of the three, simulation~5 seems most promising, where the other two lack points at low metallicity. Apparently, even with observational scatter added, short burst durations are still favored if the low metallicity objects are to be explained.

Smearing the models for simulations 8 and 9 adds some width to the predicted spread in N/O, but it does not alter our previous conclusion that differences in outflow are capable of explaining the observations. What is implied here is that with smearing, a smaller range in the wind parameter is more appropriate.

Finally, we applied scatter to the C/O points of simulation 9 and found an improved match between observation and theory. We don't show these points in Fig.~\ref{c2o} because the plot becomes too busy.

Thus, the addition of scatter to our original simulations reduces the number of parameter combinations which we can confidently rule out for explaining the distribution of objects on the N/O plateau. In particular, the possibility that massive stars can explain these objects is viable as do some types of burst models. The only scenario which clearly cannot explain the N/O plateau is one in which the galaxy age range extends below 250~Myr.

\section{SUMMARY \& CONCLUSIONS}

The problem of interpreting the morphology of the N/O plateau has been a perplexing one for over twenty years. Since the mid-90s, however, spectroscopic observations have drastically improved, and the form of the plateau has become clearer. In the most recent attempt to probe its shape, Nava et al. (2006) have recomputed N and O abundances in 68 blue compact galaxies, using published observations. Using a chi-square analysis, they have shown that most if not all of the observed scatter in N/O is due to statistical errors and is not intrinsic to the sample objects.

In the current paper, we have combined chemical evolution models with Monte Carlo simulation techniques for the first time in an attempt to assess the importance of several popular scenarios for explaining the morphology of the N/O plateau, using the results from Nava et al. as a backdrop for comparison. We have tested both the effects of time delay of N release and effective yield variations, as well as continuous and bursting star formation situations, evaluating all simulation results both before and after adding the effects of observational scatter. The former approach allowed us to study the relative impact of certain parameters and their ranges on the theoretical results, while the latter one permitted us to evaluate each simulation under more realistic restrictions.

We have reached the following conclusions.

\begin{enumerate}

\item When observational scatter is ignored, effective yield variation, as tested here using variations in selective outflow amounts, appears to be the best mechanism for explaining scatter in N/O.

\item Once scatter is accounted for, nearly all of the simulations which we performed proved to be capable of explaining the observations. In particular, ranges in star formation efficiency either by itself or coupled with system age ranges, as well as a range in selective outflow rates, produced results which proved to be consistent with the data.

\item A simulation comprising simple models over a range in SFE and galaxy age successfully fit the data when scatter was added to the output. Because nitrogen is injected immediately into the environment following star formation under simple model assumptions, this result is then consistent with the idea that massive stars produce most of the N in the universe, as already suggested by Izotov \& Thuan (1999).

\item Simulations which attempt to explain the plateau morphology by assuming a range of galaxy ages but nothing else appear to fail if the age range extends below 250~Myr. This result arises because galaxy ages which are necessary to produce objects of low metallicity are young and have not experienced complete N ejection by IMS. Therefore, they fall well below the N/O plateau. 

\end{enumerate}

Thus, the problem of the stellar origin of cosmic nitrogen remains unresolved. Indeed the data currently do not permit us to rule out enough of the studied cases to single out a clear mechanism. On the other hand, improving the S/N by reobserving BCGs using 10~meter class telescopes is likely to reduce the observational scatter and allow further testing of parameter space with respect to BCGs. Finally, the research described above has shown the capabilities of Monte Carlo techniques for interpreting observations through the use of chemical evolution models.

\acknowledgments

We are grateful to Laura Portinari for valuable correspondence and to an anonymous referee whose suggestions significantly improved the paper. We also thank Liese van~Zee
for thoughtfully reading an earlier version of the manuscript and making numerous suggestions for
improving the paper. A.N and R.B.C.H. 
are partially supported by NSF grant AST-0307118 to the University of Oklahoma, while 
J.X.P. is partially supported by NSF grant AST 03-07824 to the University of California, 
Santa Cruz.

\appendix

\section*{APPENDIX}
\section*{ NITROGEN SYNTHESIS}

Nitrogen is produced as a byproduct of the CNO bi-cycle during hydrogen burning in 
environments where the temperature exceeds 15MK. Since C is required 
as a catalyst in this process, the source of C becomes important for understanding N synthesis
and leads to the idea of secondary and primary N.
 
{\underline{Secondary Nitrogen:}} In stars with  metallicities above O/H $\approx$8.3, the dominant 
source of C is the portion which was extant in the interstellar medium at the 
time that the star formed, i.e. the C was produced by an earlier generation of stars. In this circumstance N production is sensitive to 
metallicity (since we assume that C scales directly with O) although O is not, and thus N/O is 
expected to rise with stellar metallicity. Nitrogen produced under these 
conditions is referred to as secondary nitrogen. 

{\underline{Primary Nitrogen:}} Stars also produce their own C 
from He burning via the triple alpha process, and in low metallicity stars (O/H $<$8.3) this 
source is then the dominant one, since preexisting C is scarce. Furthermore, the 
amount of C produced in this way is determined by the internal characteristics 
of the star, not its metallicity, so the nitrogen produced from it is now insensitive to stellar metallicity and is labeled primary nitrogen. In this case both N and O production are independent of metallicity, and N/O remains constant over a broad range of metallicity. The N/O 
plateau presumably arises because objects comprising it have yet to reach a 
level of metallicity where the contribution of pre-existing C 
outweighs the C produced {\it in situ} during the star's lifetime. That is to say, primary N is
dominant in this region.

\clearpage

\begin{deluxetable}{lcccccc}
\tabletypesize{\scriptsize}
\setlength{\tabcolsep}{0.07in}
\renewcommand{\arraystretch}{1.5}
\tablecolumns{7}
\tablewidth{0in}
\tablenum{1}
\tablecaption{PARAMETERS FOR MONTE CARLO SIMULATIONS}
\tablehead{
\colhead{Simulation} &
\colhead{IRA\tablenotemark{a}} &
\colhead{SFE Range (Gyr$^{-1}$)} &
\colhead{Chemical Age (Gyr)\tablenotemark{b}} & 
\colhead{Windloss Parameter} &
\colhead{Burst Separation (Gyr)} &
\colhead{Burst Duration (Gyr)}
}
\startdata
1 & yes & 0.01-0.1 & 0.02-2.0 & 0 & \nodata & \nodata \\
2 & no & 0.1 & 0.1-2.0 & 0 & \nodata & \nodata \\
3 & no & 0.005-0.09 & 2 & 0 & \nodata & \nodata \\
4 & no & 0.01-0.1 & 0.02-2.0 & 0 & \nodata & \nodata \\
5 & no & 0.075 & 1-5 & 0 & 0.5 & 0.1 \\
6 & no & 0.075 & 1-5 & 0 & 0.5 & 0.3 \\
7 & no & 0.075 & 1-5 & 0 & 0.5 & 0.4 \\
8\tablenotemark{c} & no & 0.02-0.2 & 1 & 0-0.5 & \nodata & \nodata \\ 
9\tablenotemark{d} & no & 0.002-0.02 & 10 & 0-0.5 & \nodata & \nodata  
\enddata
\tablenotetext{a} {This column indicates the status of the instantaneous recycling approximation. ``Yes'' appears if IRA is assumed, ``No'' appears if it is relaxed and finite stellar lifetimes are taken into account.}
\tablenotetext{b} {The time elapsed since the beginning of star formation.}
\tablenotetext{c}{The nitrogen yield of IMS was scaled by 0.70, while the carbon yield for massive and IMS stars was scaled by 0.30}
\tablenotetext{d}{The nitrogen yield of IMS was scaled by 0.60, while the carbon yield for massive and IMS stars was scaled by 0.30}
\end{deluxetable}

\clearpage

\end{document}